\documentclass[a4paper,11pt]{JHEP3}
\usepackage[dvips]{graphicx}
\usepackage{ifthen}
\usepackage{amsmath}
\usepackage{epsfig}
\usepackage[errorshow]{tracefnt}
\usepackage{amsfonts}
\usepackage{amsmath}
 \usepackage{amssymb}
 \newfont{\bbbold}{msbm10}
 
\def\beal{\begin{align}}
 
\def\g{\gamma}

 \def\cD{{\cal D}}

 \def\cR{{\cal R}}

 \newfont{\goth}{eufm10 scaled \magstep1}

 \def\ga{\gamma}

 \def\a{\alpha}
 \def\b{\beta}
 \def\c{\gamma}\def\C{\Gamma}
 \def\d{\delta}\def\D{\Delta}
 \def\e{\epsilon}

 \def\th{\theta}

 \def\be{\begin{equation}}\def\ee{\end{equation}}
 \def\bea{\begin{eqnarray}}\def\eea{\end{eqnarray}}
 \def\ba{\begin{array}}\def\ea{\end{array}}
 
 \def\o{\omega}
 
 \def\ua{\underline{\alpha}}
 \def\ul{\underline{\lambda}}
 \def\ub{\underline{\phantom{\alpha}}\!\!\!\beta}
 
 \def\unmu{\underline {\phantom{\alpha}}\!\!\!\mu}
 \def\unnu{\underline \nu}

 \def\una{\underline a}\def\unA{\underline A}
 \def\unb{\underline b}\def\unB{\underline B}
 \def\unc{\underline c}
 \def\und{\underline d}
 \def\une{\underline e}
 \def\unf{\underline{\phantom{e}}\!\!\!\! f}
 \def\ung{\underline g}
\def\unh{\underline h}
 \def\unm{\underline m}\def\unM{\underline M}
 \def\unn{\underline n}
 \def\unp{\underline{\phantom{a}}\!\!\! p}
 \def\unq{\underline{\phantom{a}}\!\!\! q}

 \def\str{\rm str}

 \def\3dt{\dot{3}}

\def\fg{\mathfrak{G}}

\def\cor{{\cal O}(\Psi^2,\th^5)}

\newcommand{\slsh}[1]{\displaystyle{\not} #1}

 {}

 \def\nn{\nonumber}
 \def\bd{\begin{document}}
 \def\ed{\end{document}}
 \def\ealt{\end{alignat}} 
 \def\bead{\begin{alignat}{2}}
 \def\beat{\begin{alignat}{3}}
 \def\bea{\begin{eqnarray}}
 \def\ba{\begin{array}}\def\ea{\end{array}}
 \def\eea{\end{eqnarray}}
 \def\ft#1#2{{\textstyle{{\scriptstyle #1}\over {\scriptstyle #2}}}}
 \def\fft#1#2{{#1 \over #2}}
 \newcommand{\eq}[1]{(\ref{#1})}
 \def\eqs#1#2{(\ref{#1}-\ref{#2})}
 \def\det{{\rm det\,}}
 \def\tr{{\rm tr}}\def\Tr{{\rm Tr}}
  \def\str{{\rm str}} \def\diag{{\rm diag}}
 \def\sdet{{\rm sdet}}
\def\ealt{\end{alignat}}

\title{Fivebrane instantons and Calabi-Yau fourfolds with flux}
\author{Dimitrios Tsimpis \\
Max-Planck-Institut f\"{u}r Physik --Theorie\\
F\"{o}hringer Ring 6,  80805 M\"{u}nchen, Germany\\
E-mail: \email{tsimpis@mppmu.mpg.de}}
\abstract{Using recent results on eleven-dimensional superspace, 
we compute the contribution of fivebrane instantons 
with four fermionic zeromodes in M-theory compactifications 
on Calabi-Yau fourfolds with flux. We find that no superpotential is 
generated in this case. This result is compatible with a certain flux-dependent 
modification of the arithmetic genus criterion. }
\preprint{MPP-2007-12}

\begin{document}




\section{Introduction and summary}

Recently, it has become clear that the problem of moduli stabilization may 
find its resolution in the context of flux compactifications (see 
{\it e.g.} \cite{reva, revb, revc} for recent reviews). In most recent models 
(starting with \cite{kklt}) a crucial role is played by nonperturbative effects 
which can generate a superpotential for the K\"{a}hler moduli.  
Within the context of M-theory compactifications on Calabi-Yau fourfolds, 
as was first noted in \cite{w},  the 
nonperturbative effects arise from fivebrane instantons wrapping internal 
divisors. In a dual IIB picture 
this setup is equivalent to compactifications on Calabi-Yau threefolds, with 
instantons arising from D3-branes wrapping internal divisors. 

In \cite{w} Witten showed that in the absence of flux 
a necessary condition for the generation of a superpotential  
is that the divisor which the fivebrane wraps  possesses a certain 
topological property: its arithmetic genus must be equal to one. 
When there are exactly two fermion zeromodes (corresponding to rigid isolated cycles) 
a superpotential {is} indeed generated. If more zeromodes are present, cancellations 
may occur. The lift of the arithmetic genus criterion to F-theory in general and 
IIB orientifolds in particular, was given by Robbins and Sethi in \cite{rs}.

Recently attention has been drawn to the possibility that 
the arithmetic genus criterion may be violated in the presence of 
flux \cite{ktt, saul, kall} (a discussion of the effects of flux 
was already presented in \cite{rs}). The authors of \cite{kall} defined a flux-dependent generalization of the 
arithmetic genus, $\chi_F$, 
to be discussed in more detailed in the following. $\chi_F$ is not, strictly-speaking, an index: it 
cannot be defined as the dimension of the kernel minus the dimension of the cokernel of some operator. 
At present it is not clear what should the arithmetic genus criterion be replaced 
by in the presence of fluxes. 
In particular, it is not clear whether the arithmetic genus criterion 
should simply be replaced by the condition $\chi_F=1$ or not. 
Moreover, it is conceivable that
instantons with four or more fermionic 
zero-modes contribute to the superpotential\footnote{Instantons with more 
than two zeromodes are known to contribute to higher-derivative and/or 
multi-fermion couplings \cite{bw}. Here we examine whether such instantons 
can contribute to the {\it superpotential}.}, 
as there exist higher-order fermionic terms in the 
worldvolume action of the fivebrane 
which may be used in order soak up the extra zero modes. 
Clarifying these 
issues is crucial for 
realistic model-building.

The computation of M-theory instantons goes back to the work of Becker et al \cite{bbs}. 
These techniques were further elaborated by Harvey and Moore \cite{hm} in the context 
of $G_2$ compactifications. 
The subject of fivebrane instantons in M-theory has largely remained  
unexplored, mainly due to the exotic nature of the fivebrane worldvolume theory. 
Instanton effects in heterotic M-theory have been considered in \cite{ovru, lima, angu, buch}.

Further progress beyond the computation of instantons with 
two zeromodes has been hindered by the lack of knowledge of the theta-expansions of the 
supervielbein and $C$-field in eleven-dimensional superspace.  
Recently there have been technical advances in this direction reported in \cite{t}, which applies  
the normal-coordinates approach \cite{norcor} to the case of eleven-dimensional 
superspace. 
Using this method, the expression for linear backgrounds 
was derived to all orders in $\theta$, i.e. up to and including terms 
of order $\theta^{32}$. This  
constitutes significant progress, taking to account the fact that 
previously this expansion 
was known explicitly only to order $\theta^2$ \cite{nicolai}. Results exact in the background fields 
were also presented up to and including terms of order $\theta^5$.




It is the purpose of this paper to perform 
an explicit computation in the case  of fivebrane instantons with four 
fermion zeromodes, in the context of M-theory compactifications on Calabi-Yau fourfolds  
in the presence of (normal) flux. We find that no superpotential is generated in 
this case. Therefore, our result does not rule out the possibility that in the presence of flux 
the arithmetic genus criterion should be replaced by the condition $\chi_F=1$.

As this is a somewhat technical paper, in the following subsections of the introduction
 we have tried to put it in context and to summarize 
in a self-contained way the strategy and the result of the computation.


\subsection{Review of the arithmetic genus criterion }

In \cite{w} Witten argued that M-theory compactifications on Calabi-Yau fourfolds may generate 
a nonzero superpotential in three dimensions 
through fivebrane instantons wrapping divisors of arithmetic genus one. 
We will now review his argument: 
consider a supersymmetric M-theory background of the form 
$\mathbb{R}^{1,2}\times X$, where $X$ is a Calabi-Yau fourfold\footnote{Eventually 
we will work in Euclideanized eleven-dimensional space.}. Provided a certain 
topological condition is satisfied, this is a consistent M-theory background \cite{wittflux,wittseth}.
 Compactification 
on $X$ results in an ${\cal N}=2$ theory in three dimensions (four real supercharges). 
This theory is very similar to a supersymmetric ${\cal N}=1$ theory in four dimensions, 
and we may think of it (although this is not necessary) 
as a dimensional reduction from four to three dimensions. Similarly to the case in four dimensions, 
the kinetic terms are obtained by integration over the whole superspace, whereas the Yukawa couplings and the 
mass terms are obtained by integrating over half the superspace (F-terms). Crucially, 
powerful nonrenormalization theorems prevent radiative corrections to the F-terms. 

Let us now describe the structure of the so-called `linear multiplets', 
which play a distinguished role in the discussion of \cite{w} and in the following:   
the bosonic part of a linear multiplet in four dimensions 
consists of a second-rank antisymmetric tensor 
and a real scalar. The fact that the antisymmetric 
tensor is dual in four dimensions to a scalar, can be promoted at the 
level of superfields to a duality between linear and chiral supermultiplets. 
Upon reduction to three dimensions the chiral multiplets 
give rise to chiral multiplets, whereas the linear multiplets become vector multiplets. 
In analogy to the situation in four dimensions, a 
vector in three dimensions is dual to the a scalar {\it provided there is no Chern-Simons term} 
arising from the compactification on the fourfold. 
In absence of fluxes there is indeed no Chern-Simons term which could 
obstruct the dualization, 
but this is generally no longer the case in the presence of fluxes \cite{haaca, haacb}. 
 
To be more explicit: 
upon compactification of M-theory on a Calabi-Yau fourfold,  one obtains $b_2$ vectors  
from the threeform gauge field
\beal
C=\sum_{I=1}^{b_2}A^I(x)\wedge \omega_I+\dots ~, 
\end{align}
where $x$ is a (three-dimensional) spacetime coordinate and $\{\omega_I, ~I=1,\dots b_2\}$ is a basis of 
$H^{2}(X,\mathbb{R})$, which of course coincides with $H^{1,1}(X,\mathbb{R})$ for a Calabi-Yau fourfold. 
In the absence of a 
Chern-Simons term in 
three dimensions the $A^I$s can be dualized to $b_2$ scalars, which we will call 
the `dual scalars' $\phi^I_D$, $d\phi^I_D=\star dA^I$. Note that perturbatively there are  
Peccei-Quinn symmetries whereby the dual scalars are shifted by constants; as we will see in the following, 
these continuous symmetries 
can be broken by instantons to discrete subgroups thereof. 
In addition to the $\phi_D^I$s there are $b_2$ scalars, $\phi^I$, from the deformations of the 
K\"{a}hler form  $J$,
\beal
J=\sum_{I=1}^{b_2}\phi^I(x)\omega_I
~.
\label{jexp}
\end{align}
After dualization, the bosonic fields of each vector multiplet in three 
dimensions (these are the `descendants' of the 
linear multiplets in four dimensions) consist of a pair of real scalars ($\phi^I$, $\phi^I_D$). The superpotential 
$W$ depends holomorphically on $\phi^I+i\phi^I_D$.

Following \cite{w}, we note that all terms in the superpotential depend on the vector multiplets. 
Indeed if there were any terms in the superpotential which did not depend on the vector multiplets, they 
could be computed by scaling up the metric of $X$ (since such terms would be independent of the 
K\"{a}hler class, which belongs to the vector multiplets). But in the limit where the metric is scaled up, 
M-theory reduces to supergravity  and $\mathbb{R}^{1,2}\times X$ becomes an exact solution --  
showing that there is no superpotential in this case. 

To look for instantons which may generate a superpotential, we note that the threeform gauge field 
is (magnetically) sourced by the fivebrane. Hence, a relevant instanton in three dimensions 
 is seen from the eleven-dimensional point-of-view 
as a fivebrane wrapping a six-cycle $\Sigma$ in the Calabi-Yau fourfold. In order 
for the instanton to preserve half the supersymmetry (so that it may 
generate an F-term), the cycle $\Sigma$ must be 
a holomorphic divisor. This fact is re-derived in detail in section \ref{supersymmetriccycles}, 
in the presence of normal flux. 

As can be verified explicitly, 
the contribution of the instanton includes the classical factor
\beal
\int d^2\theta_0 ~e^{-(\mathrm{Vol}_{\Sigma}+i\phi_D)}~,
\label{gras}
\end{align}
where $\mathrm{Vol}_{\Sigma}$ is the volume (in units of 
the eleven-dimensional Planck length $l_P$) of the six-cycle the fivebrane is wrapping, and $\phi_D$ is 
the linear combination of dual scalars which constitutes the superpartner of $\mathrm{Vol}_{\Sigma}$. I.e. 
the scalars ($\mathrm{Vol}_{\Sigma}$, $\phi_D$) form the real and imaginary parts of a chiral superfield, 
as is expected 
from the holomorphic property of the superpotential (which is, in its turn, a consequence of supersymmetry). 
For the generation of a superpotential, the 
fermionic terms in the fivebrane action should conspire so as to soak up all but two of the 
fermion zeromodes. The Grassmann integration in (\ref{gras}) above is the integration over the 
remaining fermionic zeromodes. As was then argued in \cite{w}, apart from the classical factor above,  
the superpotential should be independent of the K\"{a}hler class. This is because the 
dependence on $\phi_D$ is fixed by the magnetic charge of the instanton, and so the dependence on 
$\mathrm{Vol}_{\Sigma}$ is in its turn fixed by holomorphy. 

Apart from the classical factor above, 
the steepest-slope approximation of the path integral around the fivebrane instanton includes a 
one-loop determinant, which is independent of the K\"{a}hler class but depends 
holomorphically on the complex structure moduli. The one-loop result 
is in fact exact, as higher loops do not contribute to the superpotential. This can be seen as follows: 
higher loops would be proportional to positive powers of $l_P$ and would therefore 
scale as inverse powers of the volume; but, as already mentioned, apart from the classical 
factor the superpotential cannot depend on the K\"{a}hler class. 

A necessary criterion for a divisor $\Sigma$ to contribute to the superpotential is that 
its arithmetic genus $\chi$,
\beal
\chi=\sum_{p=0}^3(-1)^p h^{p,0}(\Sigma)~,
\end{align}
is equal to one. This was arrived at in \cite{w} by the following line of arguments: 
first note that, in the limit where $\Sigma$ is scaled up, the $U(1)$ 
rotations along the normal direction to $\Sigma$ inside the fourfold become an exact symmetry 
(dubbed `$W$-symmetry' in \cite{w}) of M-theory. On the other hand, in the absence of fluxes 
the worldvolume theory of the fivebrane has a one-loop $W$-anomaly equal to $\chi$. It must then 
be that the exponential in (\ref{gras}) has $W$-charge equal to $-\chi$.\footnote{Note that 
Witten's paper \cite{w} was written before the cancellation of the normal-bundle anomaly of the 
fivebrane was properly understood in \cite{hmm}. It would be interesting to 
derive this result directly using the techniques of \cite{hmm}.} 
Moreover, it is straightforward to see that 
the fermionic zeromode measure carries $W$-charge equal to one. It follows that 
a necessary condition for the generation of a superpotential is $\chi=1$; 
this is the arithmetic genus criterion.

\subsection{Caveats to the arithmetic genus criterion}
\label{caveats}

As already anticipated in \cite{w}, the arithmetic genus criterion may 
be violated in cases where the assumption of $W$-symmetry fails. This can occur if there 
are couplings of the fermions to normal 
derivatives of the background fields (i.e. normal to the divisor $\Sigma$ inside $X$). 
Indeed, in the presence of flux such couplings 
are present already in the `minimal' quadratic-fermion action $\theta\slsh\cD\theta$, 
where $\slsh\cD$ is a flux-dependent Dirac operator which we will define more precisely in the following. 
Even in the absence of flux, 
$W$-violating couplings will generally be present at higher orders in the fermions, they will 
however be suppressed in the large-volume limit. 

A further complication is the following: 
in the presence of flux, there is a Chern-Simons term in the three-dimensional 
low-energy supergravity, 
\beal
T_{IJ}d\phi^I\wedge A^J~,
\label{cs}
\end{align}
which will a priori obstruct the straightforward 
dualization of the vectors $A^I$ to scalars $\phi_D^I$ \cite{haaca,haacb}. 
One may therefore worry about the fate of holomorphy, on which the derivation of the 
 arithmetic genus criterion relied. (Recall that the holomorphic property of the superpotential allowed us to take 
the large-volume limit in which the $W$-symmetry becomes exact). The object $T_{IJ}$ which enters 
the Chern-Simons term above is a constant symmetric matrix given by
\beal
T_{IJ}&:=\frac{\partial^2 T }{\partial\phi^I\partial\phi^J}=\int_X F\wedge\omega_I\wedge\omega_J\nn\\
 T&:=\frac{1}{2}\int_X F\wedge J\wedge J=\frac{1}{2}T_{IJ}\phi^I\phi^J~,
\label{tdef}
\end{align}
where $F$ is the internal component of the fourform flux. Its quantization condition is equivalent to  
the expansion
\beal
F=\sum_{a=1}^{b_4}n^a\omega_a+\sum_{I=1}^{b_2}dA^I\wedge\omega_I~,
\end{align}
where $\{ \omega_a, ~a=1\dots b_4\}$ is a basis of $H^4(X,\mathbb{Z})$, and 
the $n^a$s are integers. 
An additional effect of the flux is the gauging of the 
the Peccei-Quinn isometries. The gauging is completely determined by the constant matrix $T_{IJ}$. 

Contrary perhaps to the na{i}ve expectation, the dualization 
of vectors to scalars can proceed more-or-less straightforwardly also 
in the case with fluxes. Let us assume for simplicity that we work in a basis 
of $H^{2}(X,\mathbb{R})$ such that $T_{IJ}$ is diagonal, and for the moment let's assume that the 
complex structure moduli are frozen. It then 
follows from the work of \cite{bhs} (which is based on general results on three-dimensional 
gauged supergravities \cite{whs}) that {\it (i)} the isometries $\phi_D^I\rightarrow \phi_D^I+\mathrm{constant}$ 
corresponding to zero eigenvalues of  $T_{IJ}$ are {\it not} gauged and {\it (ii)} if 
$\phi_D^I\rightarrow \phi_D^I+\mathrm{constant}$ is an isometry which {\it does} get gauged, the superpotential  
cannot depend on $\phi_D$ (nor can it depend on the K\"{a}hler modulus $\phi^I$, by holomorphy).\footnote{On 
the other hand, if there are additional fields 
which are charged under the gauge potential, this conclusion may be relaxed \cite{haack}. 
We thank M. Haack for pointing this out. 
In the present context, such phenomena may arise presumably in the presence of M2 branes 
\cite{gano} and will not 
be examined here. } 
This picture is consistent with the conclusions of \cite{poortomasiello} who find (in the context of 
IIA string theory) that those isometries which are gauged by the flux are protected from quantum corrections.

\subsection{The results of the present paper}

In the presence of fluxes, the scalar potential of the low-energy three-dimensional supergravity is still 
given in terms of the holomorphic superpotential $W$, but in addition will also generally depend on $T$. 
On the other hand the fermion bilinears 
\beal
\chi^I\chi^JD_ID_JW+\mathrm{c.c.}~,
\label{fbils}
\end{align}
where $D_I$ is a K\"{a}hler-covariant derivative, 
solely depend on the holomorphic superpotential, $W$, even in the presence of fluxes \cite{whs}. (
Fermion  mass terms of the form $\bar{\chi}^I\chi^J M_{IJ}$ do depend on $T$, as we will see in 
section \ref{gravitinokkreduction}).  Hence, a straightforward 
way to obtain instanton corrections to the superpotential is to compute the coupling (\ref{fbils}).

For the purpose of examining the possible generation of a superpotential by instanton effects, 
it follows from the discussion in section \ref{caveats} that 
we only need examine whether the coupling (\ref{fbils}) is generated for fermions 
$\chi^I$ which correspond to zero eigenvalues of $T_{IJ}$ (we may consider a 
basis where $T_{IJ}$ is diagonal, for simplicity). 
Hence, we may assume that the K\"{a}hler moduli corresponding 
to nonzero eigenvalues of $T_{IJ}$ are frozen to zero\footnote{Examples of fourfolds for which there are 
choices of fourform flux such that $T_{IJ}$ vanishes identically, were examined in \cite{mayra}.}. 
In other words we can assume, as follows 
from (\ref{jexp},\ref{tdef}), 
that we are in the region of the K\"{a}hler moduli space where:
\beal
\int_{X}F\wedge J\wedge\omega_I =0; ~~~~~I=1\dots b_2~.
\label{wderiv}
\end{align}
If no such region exists, {\it i.e.} if $T_{IJ}$ has no zero eigenvalues, all isometries are gauged 
and there can be no superpotential dependence on the K\"{a}hler moduli: 
the superpotential is protected against instanton contributions. Moreover, condition (\ref{wderiv}) 
implies that 
\beal
\omega_I\lrcorner F=0~, 
\label{dkn}
\end{align}
for all $\omega_I$s corresponding to zero eigenvalues of $T_{IJ}$. 
This observation simplifies somewhat the rather tedious computational task of this paper. 
In particular, we may assume we are in the region of the K\"{a}hler moduli space where the 
flux is primitive: $J\lrcorner F=0$. Furthermore, for the purposes of the present  
computation we may assume that the complex structure moduli are frozen to values 
such that the internal fourform flux is of type (2,2). These are exactly the 
conditions which ensure that 
{\it the flux is compatible with supersymmetry}, 
as we will see in detail in section \ref{mtheoryonfourfolds}.

Despite the fact that certain conceptual subtleties remain, there are clear rules for instanton 
computations in M-theory first put forward in \cite{bbs} and 
subsequently  elucidated in \cite{hm}. We will schematically describe the procedure here, relegating 
the details to the main body of the paper. In order to compute the instanton 
contribution to the coupling (\ref{fbils}), one first decomposes the eleven-dimensional gravitino  
in terms of three-dimensional  fermions $\chi^I$,
\beal
\Psi_m=\chi^I\otimes\Omega_{I,m}\xi~,
\label{kkgravit}
\end{align}
where $\xi$ is the covariantly constant spinor of the Calabi-Yau fourfold\footnote{In the 
presence of flux, the internal space becomes a warped Calabi-Yau. As we will see, however, the 
effect of the warp factor can be ignored at leading order in the large-volume expansion.} 
and $\Omega_I$ is a one-form on $X$ valued in the Clifford algebra $Cl(TX)$. Next, from the 
fivebrane action one reads off the coupling of the eleven-dimensional 
gravitino to the fivebrane worldvolume fermion $\theta$, schematically:
\beal
V=\sum_{n} c_n\Psi\theta^{2n+1} ~,
\end{align}
for some, possibly flux-dependent, `coefficients' $c_n$. 
The coupling $V$ is the `gravitino vertex operator'. Finally, to read off the 
coefficient $D_ID_JW$ in (\ref{fbils}) one evaluates the correlator $\langle VV\rangle$ 
in the worldvolume theory of the fivebrane.

Note that the worldvolume fermions are valued in the normal bundle to the fivebrane, which is 
the sum of $T\mathbb{R}^3$ (after passing to Euclidean signature) 
and the normal bundle to the divisor inside the fourfold. 
Thus, each worldvolume 
fermion should be thought of as  tensored with a two-component spinor of $Spin(3)$. 
 The main result of the present paper is that {\it instantons with exactly four fermionic zeromodes 
do not contribute to the superpotential.} In deriving this result we have made the 
simplifying assumption that both the curvature of the worldvolume self-dual tensor 
as well as the  pull-back of the threeform flux onto the 
worldvolume vanish. This is what we call the condition of `normal flux'.

One major technical difficulty with the present computation is the explicit expansion of the 
fivebrane action in terms of the worldvolume fermion, the so-called 
`theta-expansion'. This, in its turn, 
stems from the theta-expansion of the eleven-dimensional background superfields 
on which the fivebrane action depends. Until recently, this expansion had only been fully worked 
out to quadratic order in the fermions. The present computation is now possible 
thanks to the recent results of \cite{t} in which, among other things, the 
theta-expansion of the eleven-dimensional superfields was computed explicitly 
to fifth order in the fermions.

We should at this point elaborate on what we mean by `the fivebrane action'. The fivebrane dynamics 
was given in terms of covariant field equations in  \cite{howea, howeb}. For the application 
we are interested in, however, one needs to work with an action. As is well known, the 
worldvolume theory of the fivebrane contains a self dual antisymmetric tensor which 
renders the formulation of an action problematic. A covariant supersymmetric action 
for the fivebrane can be constructed 
with the help of an auxiliary scalar \cite{pst}. Alternatively, the auxiliary field can 
be eliminated at the expense of explicitly breaking Lorentz invariance \cite{schw}. The equivalence of all 
different formulations was shown in \cite{equi}. Here we will 
use the covariant action of \cite{pst}.

An important cautionary remark is in order. 
In \cite{wfive} Witten pointed out that a useful way to define the action of a 
self-dual field is in terms of 
a Chern-Simons theory in one dimension higher. This definition, for spacetime 
dimensions higher than two, involves a suitable 
generalization of the notion of spin structure -- on 
a choice of which the self-dual action depends. These issues have  been recently
clarified by Belov and Moore \cite{beloa, belob}. Unfortunately, the action of \cite{pst} does not take 
these topological aspects into account; it is however at present our only available 
(covariant) { supersymmetric action} for the fivebrane.

\subsection{Outline}

We now give a detailed plan of the rest of the paper. Section \ref{thetaexpansions} 
relies on \cite{t} treating 
 the theta-expansion of the various superfields of the eleven-dimensional background, 
with the aim of applying it to the worldvolume theory of the fivebrane. The theta 
expansion of the sixform potential was not considered in \cite{t}, and this is addressed in 
section \ref{thetaexpansions1}. The worldvolume theory of the fivebrane 
is considered in section \ref{pst} in the framework of the covariant action of \cite{pst}. 
Eventually we make the simplifying assumption that the flux is `normal', {\it i.e.} 
that both the field-strength of the worldvolume antisymmetric tensor and the pull-back of the background 
threeform flux onto 
the fivebrane worldvolume, vanish. 
The main result of this section is the form of the gravitino vertex operator 
in the case of normal flux, equation (\ref{grv}).

Section \ref{mtheoryonfourfolds} 
considers M-theory backgrounds of the form of a warp product 
$\mathbb{R}^{1,2}\times_{w}X$, where $X$ is a Calabi-Yau fourfold. 
(Eventually we Wick-rotate to Euclidean signature and take the large-volume limit 
in which the warp factor becomes trivial). 
Requiring 
${\cal N}=2$ supersymmetry in three dimensions (four real supercharges) implies certain 
restrictions on the fourform flux, equation (\ref{gform}). 
Next we consider fivebrane instantons such that the worldvolume 
wraps a six-cycle ${\Sigma} \subset X$ and we assume that $X$ can be 
thought of as the total space 
of the normal bundle of ${\Sigma}$ inside $X$. As discussed in the introduction, 
this approximation becomes 
more accurate as the size of ${\Sigma}$ is scaled up. 
Imposing the normal-flux condition,  
the form of the background flux simplifies further, equations 
(\ref{fffn}, \ref{nfff}).

In section \ref{supersymmetriccycles} 
we show that, in the case of normal flux, demanding that the instanton 
preserve one-half the supersymmetry of the background 
implies   that ${\Sigma}$ is an 
 (anti)holomorphic cycle. Section \ref{zeromodes} 
treats the worldvolume fermion zeromodes of 
the flux-dependent Dirac operator, 
equation (\ref{dirac}). After decomposing the background fermion 
in terms of forms on the fivebrane, we 
derive the explicit expression of the fermion zeromodes (\ref{zm}). 
This result agrees with the analysis of \cite{saul, kall}, 
in the case of normal flux and 
provided the warp factor is trivial. This can be consistently taken 
to be the case in the large-volume limit, as explained in 
section \ref{mtheoryonfourfolds}.

In section \ref{instantoncontributions} we finally come to the main subject of the paper, 
the instanton contributions 
to the superpotential. 
Section \ref{gravitinokkreduction} discusses the Kaluza-Klein Ansatz for the 
gravitino, equation (\ref{kkgr}). 
%
%
Next, the Kaluza-Klein ans\"{a}tze 
for the gravitino as well as for the fermion zeromodes are substituted into 
the expression (\ref{grv}) for the gravitino vertex operator. 
The result of the fermion zeromode integration in the case of two zeromodes 
is briefly discussed in section \ref{ofrzm}. 
In section \ref{frzm} it is shown that in the case 
of four fermion zeromodes 
the result of the zeromode integration is zero. I.e. in this case the 
instanton contribution to the superpotential vanishes. 

The appendices contain several useful technical details. For quick reference, 
 we have also included an  index of our conventions and notation in 
section \ref{notation/conventions}.

\section{Theta-expansions}
\label{thetaexpansions}

This section examines the theta-expansions of the various eleven-dimensional superfields. Except 
for the expansion of the sixform which is given in section \ref{thetaexpansions1}, these 
were treated in reference \cite{t} to which the reader is referred for 
further details. 
For reasons which are explained below (\ref{grv}), 
for our purposes we will not need the explicit form of the $\Psi^2$ contact terms. 
It also suffices to keep terms up to and including order $\th^3$. Also note that 
we are using standard superembedding notation, whereby target-space indices are underlined. 
Further explanation of the notation can be found in appendix \ref{notation/conventions}.

\subsection{Vielbein and threeform } 

Using the formul{\ae} in \cite{t}, to which the interested reader is referred 
for further details, we find
\bead
E_{m}{}^{\una}&=e_{m}{}^{\una}
-\frac{i}{2}(\cD_m\th\C^{\una}\th)
+\frac{1}{24}(\cD_m\th\mathfrak{G}\C^{\una}\th)
+\frac{1}{24}(\th\cR_{\unn\unp}{\cal I}_m{}^{\unn\unp}\C^{\una}\th)\nn\\
&-i(\Psi_m\C^{\una}\th)+\frac{1}{6}(\Psi_m\fg\C^{\una}\th)
+\frac{1}{6}(\Psi_{\unn\unp}{\cal I}_m{}^{\unn\unp}\C^{\una}\th)
+{\cal O}(\Psi^2, \th^5)~,
\label{v}
\end{alignat}
\vfill\break
where
\bead
(\fg)_{\ua}{}^{\ub}&:=\frac{1}{576}\Big\{
(\th\C^{\una\unb\unc\und\une\unf})_{\ua}(\th\C_{\une\unf})^{\ub}
-2(\th\C_{\une})_{\ua}(\th\C^{\una\unb\unc\und\une})^{\ub}
-16(\th\C^{\una})_{\ua}(\th\C^{\unb\unc\und})^{\ub}\nn\\
&+24(\th\C^{\una\unb})_{\ua}(\th\C^{\unc\und})^{\ub}
\Big\}G_{\una\unb\unc\und}~,
\label{fgdef}
\end{alignat}
\bead
({\cal I}_{m}{}^{\une\unf})_{\ua}{}^{\ub}&:=-\frac{1}{48}\Big\{
(\th\C_{\una\unb})_{\ua}(\th\C_m{}^{\una\unb\une\unf})^{\ub}
+4(\th\C_{m\una})_{\ua}(\th\C^{\una\une\unf})^{\ub}
-4(\th\C_{\una\unb})_{\ua}(\th\C^{\una\unb\une})^{\ub}e_m{}^{\unf}  \nn\\
&+6(\th\C_{m})_{\ua}(\th\C^{\une\unf})^{\ub}
-12(\th\C_{\una})_{\ua}(\th\C^{\una\une})^{\ub}e_m{}^{\unf}
\Big\}
~.
\end{alignat}
Using (\ref{v}) we find for the Green-Schwarz metric
\bead
g_{mn}&=G_{mn}-\frac{1}{4}(\cD_m\th\C^{\una}\th)(\cD_n\th\C_{\una}\th)
-i(\cD_{(m}\th\C_{n)}\th)+\frac{1}{12}(\cD_{(m}\th\fg\C_{n)}\th)\nn\\
&+\frac{1}{12}(\th\cR_{\unp\unq}{\cal I}_{(m}{}^{\unp\unq}\C_{n)}\th)   -2i(\Psi_{(m}\C_{n)}\th)
+\frac{1}{3}(\Psi_{(m}\fg\C_{n)}\th) \nn\\
&+\frac{1}{3}(\Psi_{\unp\unq}{\cal I}_{(m}{}^{\unp\unq}\C_{n)}\th)
-(\Psi_{(m}\C^{\una}\th)(\cD_{n)}\th\C_{\una}\th)+{\cal O}(\Psi^2,\th^5)
~.
\end{alignat}
Similarly, for the pull-back of the three-form we find
\bead
C_{mnp}&=
c_{mnp}-\frac{3i}{2}(\cD_{[m}\th\C_{np]}\th)+\frac{1}{8}(\cD_{[m}\th\fg\C_{np]}\th)
+\frac{1}{8}(\th\cR_{\unp\unq}{\cal I}_{[m}{}^{\unp\unq}\C_{np]}\th)
\nn\\
&-\frac{3}{4}(\cD_{[m}\th\C_{n}{}^{\una}\th)(\cD_{p]}\th\C_{\una}\th)
-3i(\Psi_{[m}\C_{np]}\th)-(\Psi_{[m}\C_{n}{}^{\una}\th)(\cD_{p]}\th\C_{\una}\th )\nn\\
&-2(\Psi_{[m}\C^{\una}\th)(\cD_{n}\th\C_{p]}{}_{\una}\th)
+\frac{1}{2}(\Psi_{[m}\fg\C_{np]}\th)+\frac{1}{2}(\Psi_{\unn\unq}{\cal I}_{[m}{}^{\unn\unq}\C_{np]}\th)
+\cor~.
\end{alignat}
%


\subsection{Sixform} 
\label{thetaexpansions1}

The $\th$-expansion for $C_6$ was not given in \cite{t}, but the same methods 
can be applied in this case. First we note that the $C_6$-field satisfies
\beal
7\partial_{[\unM_1}C_{\unM_2\dots \unM_7\}}=G_{\unM_1\dots \unM_7}.
\label{bianchi}
\end{align}
Up to a gauge choice, 
the following is a solution of
the Bianchi identity (\ref{bianchi}) at each order 
in the $\theta$ expansion:
\bead
C^{(0)}_{\unmu_1\dots\unmu_6}=
C^{(0)}_{\unmu_1\dots\unmu_5 \unm_1}&=\dots 
C^{(0)}_{\unmu_1 \unm_1\dots \unm_5}=0~,\nn\\
7\partial_{[\unm_1}C^{(0)}_{\unm_2\dots \unm_7]}&=G^{(0)}_{\unm_1\dots \unm_7}
\label{cexpa}
\end{alignat}
\vfill\break
and
\bead
C^{(n+1)}_{\unmu_1\dots\unmu_6}&=\frac{1}{n+7}~\theta^{\ul} G^{(n)}_{{\ul}\unmu_1\dots\unmu_6}\nn\\
C^{(n+1)}_{\unmu_1\dots\unmu_5 \unm_1}&=\frac{1}{n+6}~\theta^{\ul} G^{(n)}_{{\ul}\unmu_1\dots\unmu_5 \unm_1}\nn\\
C^{(n+1)}_{\unmu_1\dots\unmu_4 \unm_1\unm_2}&=\frac{1}{n+5}~\theta^{\ul} G^{(n)}_{{\ul}\unmu_1\dots\unmu_4 \unm_1\unm_2}\nn\\
C^{(n+1)}_{\unmu_1\unmu_2\unmu_3 \unm_1\unm_2 \unm_3}&=\frac{1}{n+4}~\theta^{\ul} G^{(n)}_{{\ul}\unmu_1\unmu_2\unmu_3 \unm_1\unm_2 \unm_3}\nn\\
C^{(n+1)}_{\unmu_1\unmu_2 \unm_1\dots \unm_4}&=\frac{1}{n+3}~\theta^{\ul} G^{(n)}_{{\ul}\unmu_1\unmu_2 \unm_1\dots \unm_4}\nn\\
C^{(n+1)}_{\unmu \unm_1\dots \unm_5}&=\frac{1}{n+2}~\theta^{\ul} G^{(n)}_{{\ul}\unmu \unm_1\dots \unm_5}\nn\\
C^{(n+1)}_{\unm_1\dots \unm_6}&=\frac{1}{n+1}~\theta^{\ul} G^{(n)}_{{\ul} \unm_1\dots \unm_6}
~, ~~~~~n\geq 0~.
\label{cexp}
\end{alignat}
Using the fact that
\be
G_{\una_1\dots \una_5{\ua}_1{\ua}_2}=-i(\C_{\una_1\dots \una_5})_{{\ua}_1{\ua}_2}
~,
\end{equation}
we find for the right-hand sides of the equations (\ref{cexp}),
\bead
\theta^{\ul} G_{{\ul}\unmu_1\dots\unmu_6}&=  6iE_{(\unmu_1}{}^{{\una}_1} \dots E_{\unmu_5}{}^{{\una}_5} E_{\unmu_6)}{}^{\ua}
(\C_{{\una}_1\dots {\una}_5}\theta)_{\ua}  \nn\\
\theta^{\ul} G_{{\ul}\unmu_1\dots\unmu_5 \unm}&=  
-5i E_m{}^{{\una}_1} E_{(\unmu_1}{}^{{\una}_2} \dots E_{\unmu_4}{}^{{\una}_5} E_{\unmu_5)}{}^{\ua}
(\C_{{\una}_1\dots {\una}_5}\theta)_{\ua}   \nn\\
&~~~+i  E_{\unmu_1}{}^{{\una}_1} \dots E_{\unmu_5}{}^{{\una}_5} E_{\unm}{}^{\ua}
(\C_{{\una}_1\dots {\una}_5}\theta)_{\ua}   
\nn\\
\theta^{\ul} G_{{\ul}\unmu_1\dots\unmu_4 \unm_1\unm_2}&=  
4i E_{\unm_1}{}^{{\una}_1} E_{\unm_2}{}^{{\una}_2} E_{(\unmu_1}{}^{{\una}_3} E_{\unmu_2}{}^{{\una}_4} E_{\unmu_3}{}^{{\una}_5} E_{\unmu_4)}{}^{\ua}
(\C_{{\una}_1\dots {\una}_5}\theta)_{\ua}   \nn\\
&~~~+2i  E_{\unmu_1}{}^{{\una}_1} \dots E_{\unmu_4}{}^{{\una}_4} E_{[\unm_1}{}^{{\una}_5} E_{\unm_2]}{}^{\ua}
(\C_{{\una}_1\dots {\una}_5}\theta)_{\ua}   
 \nn\\
\theta^{\ul} G_{{\ul}\unmu_1\unmu_2\unmu_3 \unm_1\unm_2 \unm_3}&=    
-3i E_{\unm_1}{}^{{\una}_1} E_{\unm_2}{}^{{\una}_2} E_{\unm_3}{}^{{\una}_3} E_{(\unmu_1}{}^{{\una}_4} E_{\unmu_2}{}^{{\una}_5} E_{\unmu_3)}{}^{\ua}
(\C_{{\una}_1\dots {\una}_5}\theta)_{\ua}   \nn\\
&~~~+3i  E_{\unmu_1}{}^{{\una}_1}E_{\unmu_2}{}^{{\una}_2} E_{\unmu_3}{}^{{\una}_3} E_{[\unm_1}{}^{{\una}_4}E_{\unm_2}{}^{{\una}_5} E_{\unm_3]}{}^{\ua}
(\C_{{\una}_1\dots {\una}_5}\theta)_{\ua}    
\nn\\
\theta^{\ul} G_{{\ul}\unmu_1\unmu_2 \unm_1\dots \unm_4}&=     
+2i E_{\unm_1}{}^{{\una}_1} \dots E_{\unm_4}{}^{{\una}_4} E_{(\unmu_1}{}^{{\una}_5} E_{\unmu_2)}{}^{\ua}
(\C_{{\una}_1\dots {\una}_5}\theta)_{\ua}   \nn\\
&~~~+4i  E_{\unmu_1}{}^{{\una}_1}E_{\unmu_2}{}^{{\una}_2} E_{[\unm_1}{}^{{\una}_3}E_{\unm_2}{}^{{\una}_4}E_{\unm_3}{}^{{\una}_5} E_{\unm_4]}{}^{\ua}
(\C_{{\una}_1\dots {\una}_5}\theta)_{\ua}    
  \nn\\
\theta^{\ul} G_{{\ul}\unmu \unm_1\dots \unm_5}&=     
-i E_{\unm_1}{}^{{\una}_1} \dots E_{\unm_5}{}^{{\una}_5} E_{\unmu}{}^{\ua}
(\C_{{\una}_1\dots {\una}_5}\theta)_{\ua}   \nn\\
&~~~+5i  E_{\unmu}{}^{{\una}_1} E_{[\unm_1}{}^{{\una}_2}\dots E_{\unm_4}{}^{{\una}_5} E_{\unm_5]}{}^{\ua}
(\C_{{\una}_1\dots {\una}_5}\theta)_{\ua}       \nn\\
\theta^{\ul} G_{{\ul} \unm_1\dots \unm_6}&= 6i E_{[\unm_1}{}^{{\una}_1}\dots E_{\unm_5}{}^{{\una}_5} E_{\unm_6]}{}^{\ua}
(\C_{{\una}_1\dots {\una}_5}\theta)_{\ua}    
~.
\label{cexpr}
\end{alignat}
In the following we will only need the part $\Delta C_6$ of $C_6$ which is linear in 
the gravitino. Plugging the expressions for the vielbein components 
given in \cite{t} into (\ref{cexpr}) we obtain
\bead
\Delta C_{m_1\dots m_6}=
&-6i(\Psi_{[m_1}\C_{m_2\dots m_6]}\th) 
+10(\Psi_{[m_1}\C^{\una}\th)(\cD_{m_2}\th\C_{m_3\dots m_6]\una}\th)\nn\\
&+(\Psi_{[m_1}\fg\C_{m_2\dots m_6]}\th)
+(\Psi_{\unp\unq}{\cal I}_{[m_1}{}^{\unp\unq}\C_{m_2\dots m_6]}\th)\nn\\
&-5(\Psi_{[m_1}\C_{m_2\dots m_5 \una}\th)(\cD_{m_6]}\th\C^{\una}\th)
+\cor~.
\label{dc6}
\end{alignat}


\section{Fivebrane action}
\label{pst}

We are now ready to consider the application of the theta-expansion 
discussed in the previous section to the case of the fivebrane 
worldvolume action. 
As already mentioned in the introduction, we will adopt the covariant framework of \cite{pst} to which 
the reader is referred for more details. The main result of this 
section is the gravitino vertex operator, equation (\ref{grv}) below. 
To improve the presentation, we have relegated
 the details of the derivation to appendix \ref{pstapp}.

The fivebrane action is of the form
\bead
S=S_1+S_2+S_3~,
\end{alignat}
where
\bead
S_1&:=T_{M5}\int_{{\Sigma}} d^6x\sqrt{-det(g_{mn}+i\widetilde{H}_{mn} ) }\nn\\
S_2&:=T_{M5}\int_{{\Sigma}} d^6x\sqrt{-g}~ \frac{1}{4}{\widetilde{H}}_{mn} {H}^{mn}\nn\\
S_3&:=T_{M5}\int_{{\Sigma}} \Big(C_6+\frac{1}{2}F_3\wedge C_3 \Big)~
\label{action}
\end{alignat}
and $T_{M5}\sim l_P^{-6}$ is the fivebrane tension. 
Moreover, we have made the following definitions
\bead
H_{mnp}&:=F_{mnp}-C_{mnp}\nn\\
H_{mn}&:=H_{mnp}v^p\nn\\
\widetilde{H}^{mn}&:=\frac{1}{6\sqrt{-g}}\e^{mnpqrs}v_pH_{qrs}\nn\\
v_p&:=\frac{\partial_p a}{\sqrt{-g^{mn}\partial_ma\partial_n a}}~,
\end{alignat}
where $F_{mnp}$ is 
the field-strength of the world-volume chiral two-form and $a$ is an auxiliary 
world-volume scalar. It follows from the above definitions that
\bead
det(\d_{m}{}^{n}+i\widetilde{H}_{m}{}^{n} )=
1+\frac{1}{2}tr\widetilde{H}^2+\frac{1}{8}(tr\widetilde{H}^2)^2-\frac{1}{4}tr\widetilde{H}^4~.
\end{alignat}
%


\subsection{The gravitino vertex operator}
\label{grvsec}

In the case of normal flux, {\em i.e.} when the world-volume two-form tensor is flat 
($F_{mnp}=0$) and  
the pull-back of the three-form potential onto the fivebrane vanishes 
($c_{mnp}=0$), the expression for the gravitino 
vertex operator simplifies considerably. Skipping the details 
of the derivation, which can be found in appendix \ref{pstapp}, the 
final result reads:
\begin{center}
\fbox{\parbox{14.5cm}{
%
%
\bead
V=  T_{M5}\int_{{\Sigma}}d^6x\sqrt{-G}~
\Big\{
2(\Psi_m\C^m\theta)+i(\Psi_{\unm} V^{(2)\unm})
&+\frac{i}{3}(\Psi_m\fg\C^m\theta)
\nn\\
&+\frac{i}{3}(\Psi_{\unp\unq}{\cal I}_m{}^{\unp\unq}\C^m\theta)
+{\cal O}(\Psi^2, \theta^5)
\Big\}
~.\nn
\end{alignat}
%
%
}}
\end{center}
\beal\label{grv}\end{align}
We can now see why the $\Psi^2$ contact-terms can be neglected. As is easy to verify, 
$\Psi^2$ terms first appear in the $\th$-expansion at order $\th^4$. 
Consequently, a single vertex-operator insertion $V_{\Psi^2}$ is needed 
to saturate the four fermion zeromodes --which is the 
case examined here. A single insertion, however, is proportional 
to $T_{M5}$ and is of order ${\cal O}(l_P^6)$ relative to  two 
vertex-operator insertions: the latter give a contribution 
proportional to $T_{M5}^2$. Clearly, this analysis is valid 
provided the `radius' of the six-cycle is much larger than the Planck length, 
${\rm Vol}_{\Sigma}>>l^6_P$.

As was shown in the case of the M-theory membrane \cite{hklt} and is 
also expected in the case of the fivebrane \cite{dk},  the first higher-order 
correction to the world-volume action occurs at order $l_P^4$. Hence it would be 
inconsistent to include contact terms without considering the higher-order 
derivative corrections to the world-volume action. Moreover, at order 
$l_P^6$ (eight derivatives) there are higher-order curvature corrections to the background supergravity 
action\footnote{The eleven-dimensional supergravity 
admits a supersymmetric deformation at order $l_P^3$ (five derivatives) \cite{ttt}. On a 
topologically-nontrivial spacetime $M$ such that $p_1(M)\neq 0$,  
this deformation can be removed by a $C$-field redefinition, at the cost of shifting the 
quantization condition of the fourform fieldstrentgh.} which,  
as was explained in \cite{t}, modify the $\th$-expansion of all superfields.


\subsection{Quadratic fermion terms}

It follows from the preceding sections that 
in a bosonic background ($\Psi_{\unm}^{\ua}=0$) the part of the Lagrangian quadratic in $\th$ 
(this is the analogue of equations (38), (39) of \cite{tt}) is given by
\bead
{\cal L}^{(quad)}&=  \frac{i}{2}\sqrt{det(A_i{}^j)}(A^{-1})^{(mn)}(\th\C_{m}\cD_n\th)\nn\\
&-\frac{\e^{lpqrs}{}_m}{6\sqrt{-G}}\sqrt{det(A_i{}^j)}(A^{-1})^{[mn]}
(\th\C_{(n}\cD_{l)}\th) a_p (F_{qrs}-c_{qrs})\nn\\
&-\frac{\e^{klpqrs}}{24\sqrt{-G}}\sqrt{det(A_i{}^j)}(A^{-1})_{kl}
a_p \nn\\
&~~~~~~~~~~~~~~~~~\times\Big\{
(F_{qrs}-c_{qrs})\Big[ a^ma^n(\th\C_m\cD_n\th) +(\th\C^{m}\cD_m\th)
\Big] +3(\th\C_{qr}\cD_s\th) \Big\} \nn\\
&-\frac{i\e^{klpqrs}}{24\sqrt{-G}}
a_ka^m(F_{lpq}-c_{lpq})\nn\\
&~~~~~~~~~~~~~~~~~\times\Big\{(F_{rst}-c_{rst})
\Big[a^ta^n (\th\C_n\cD_m\th)+\frac{1}{2}(\th\C^t\cD_m\th) \Big]
+\frac{1}{2}(\th\C_{rs}\cD_m\th)
\Big\} \nn\\
&-\frac{i\e^{klpqrs}}{48\sqrt{-G}}
a_ka^n (F_{lpq}-c_{lpq}) (F_{rs}{}^{t}-c_{rs}{}^{t})
(\th\C_n\cD_t\th)\nn\\
&-\frac{i\e^{klpqrs}}{2\times 5!\sqrt{-G}}
\Big\{  15  a^ta_k(F_{lpt}-c_{lpt})(\th\C_{qr}\cD_s\th) 
-10 a^ta_k(F_{lpq}-c_{lpq})(\th\C_{rt}\cD_s\th)\nn\\
&~~~~~~~~~~~~~~~~~~~~~~~~~~~~~~~~~~~~~~~~~~~~~~~~~~~~~~~~~~~~
- 5F_{klp}(\th\C_{qr}\cD_s\th)  -(\th\C_{klpqr}\cD_s\th) 
\Big\}
~.
\end{alignat}
Note that ${\cal L}^{(quad)}$ is related to $V_{~~~\ua}^{(1)\unm}$ in a simple way.

\subsection*{Normal  flux}

In this case the part of the Lagrangian quadratic in the fermions  
simplifies to
\bead
{\cal L}^{(quad)}&=  \frac{i}{2}\Big\{ (\th\C^{m}\cD_m\th)
+\frac{\e^{klpqrs}}{ 5!\sqrt{-G}}
(\th\C_{klpqr}\cD_s\th) \Big\}
~.
\end{alignat}
After Wick-rotating we obtain
\bead
{\cal L}^{(quad)}&= -(\th\C^{m}\cD_m\th)
~,
\label{ityu}
\end{alignat}
where we have taken (\ref{gammahodge}) into account, and 
we have noted that after gauge-fixing the physical fermion modes satisfy $P^+\th=\th$.


\section{Supersymmetric cycles}

This section is devoted to the analysis of the conditions for a 
supersymmetric six-cycle, and the derivation 
of the worldvolume fermionic zeromodes  in the presence of (normal) flux.


\subsection{M-theory on fourfolds}
\label{mtheoryonfourfolds}

We start by reviewing M-theory on 
a Calabi-Yau fourfold with flux. Let the eleven-dimensional metric be of the form
\beal
ds^2=\Delta^{-1}ds_3^2+\Delta^{1/2}ds^2_8~,
\end{align}
where $ds_3^2$ is the metric of three-dimensional Minkowski space, 
$\Delta$ is a warp factor, and 
$ds^2_8$ 
is the metric on $X$. Let us also decompose the eleven-dimensional Majorana-Weyl supersymmetry 
parameter 
$\eta$ in terms of a real anticommuting spinor $\epsilon$ along the three-dimensional  
Minkowski space, and a real chiral spinor $\xi$ on $X$:
\beal
\eta=\Delta^{-1/4}\epsilon\otimes\xi~.
\label{ansa}
\end{align}
As was first shown in \cite{bb}, the requirement of 
${\cal N}=1$ supersymmetry in three dimensions (two real supercharges) leads 
to the condition 
\beal
\nabla_m\xi=0~,
\end{align}
i.e. the `internal' spinor is covariantly constant with respect to the connection associated 
with the metric $g_{mn}$ on $X$. Under the Ansatz (\ref{ansa}), requiring 
${\cal N}=2$ supersymmetry in three dimensions implies the existence of two 
real covariantly-constant spinors $\xi_{1,2}$ of the same chirality. 
It follows that $X$ is a Calabi-Yau four-fold. In the following 
we shall combine $\xi_{1,2}$ 
into a complex chiral spinor on $X$, $\xi:=\xi_1+i\xi_2$. An antiholomorphic 
$(0,4)$ fourform $\Omega$ and a complex structure $J$ on $X$ can be constructed as 
bilinears of $\xi$, as is discussed in detail in appendix \ref{sus}.
Moreover, 
supersymmetry imposes 
the following conditions on the components of the fourform field-strength:
\beal
G={\rm Vol}_3\wedge d\Delta^{-3/2}+F~,
\label{gform}
\end{align}
where $F$ is a fourform on $X$ which is purely $(2,2)$ and traceless, 
$J\lrcorner F=0$, with respect to the 
complex structure $J$ on $X$. We have 
denoted by  ${\rm Vol}_3$ the volume element of the three-dimensional 
Minkowski space. 
Finally, the warp factor is constrained by the Bianchi identities to satisfy 
\beal
d\star d~{\rm log}\Delta=\frac{1}{3}F\wedge F-\frac{2}{3}(2\pi)^4\beta X_8~,
\label{x8}
\end{align}
where $\beta$ is a constant of 
order $l_P^6$, and  the Hodge star is with respect to the metric on $X$. 
The second term on the right-hand side of the equation above is a higher-order correction 
related to the fivebrane anomaly. In general there will be other corrections of the same order 
which should also be taken into account. However, it can be argued that 
in the large-radius approximation it is consistent 
to only take the above correction into account (see \cite{pvw}, for example).

In the large-volume limit $g^{CY}=tg_{0}^{CY}+\dots$, $t\rightarrow\infty$, the two terms on the 
right-hand side of (\ref{x8}) scale like $t^{-3}$ relative to the left-hand side and can be 
neglected. It is therefore consistent to take the warp factor to be trivial, $\Delta=1$ \cite{beck}. 
We will henceforth assume this to be the case. In particular, it follows from (\ref{gform}) that 
the fourform's only nonzero components are along the Calabi-Yau fourfold. 
Note that the integrated version 
of equation (\ref{x8}), 
\beal
\int_X F\wedge F+\frac{\beta}{12} ~\chi(X)=0~,
\end{align}
is the tadpole cancellation condition.

Finally, note that the normal flux condition, together with the constraints of supersymmetry 
on the fourform flux explained in section 
\ref{mtheoryonfourfolds}, imply that $F$ is of the form
\beal
F_{mnpq}=4\widetilde{F}_{[mnp}K_{q]}+4\widetilde{F}^*_{[mnp}K^*_{q]}~,
\label{fffn}
\end{align}
where $\widetilde{F}$ obeys 
\beal
{J}\lrcorner \widetilde{F}=0; ~~~~~
\iota_K\widetilde{F}=\iota_{K^*}\widetilde{F}=0~
\label{nfff}
\end{align}
and $K$ is a complex vector field normal to the 
fivebrane worldvolume, see eq. (\ref{see}) below.

The above results can be extended to include more general fluxes 
\cite{ms, teight}. In this case the internal 
manifold generally ceases to be Calabi-Yau.


\subsection{Supersymmetric cycles}
\label{supersymmetriccycles}

Consider a bosonic superembedding of the fivebrane ($X^{\unm}(\sigma), ~\th^{\unmu}(\sigma)=0)$ 
in a bosonic background $(\Psi_ {\unm}{}^{\ua}=0)$, where $\sigma^m$ is 
the coordinate on the fivebrane worldvolume. 
The fivebrane action is invariant under superdiffeomorphisms 
\bead
\d_\zeta Z^{\unM}=\zeta^{\unA} E_{\unA}{}^{\unM}
\label{a}
\end{alignat}
such that 
\bead
{\cal L}_{\zeta}E_{\unM}{}^{\unA}= -(\partial_{\unM}+\Omega_{\unM\unB}{}^{\unA})\zeta^{\unB}
-\zeta^{\unB}T_{\unB\unM}{}^{\unA} =0~.
\label{ui}
\end{alignat}
%
This can be seen by first noting that
\bead
{\cal L}_{\zeta}C_3=d(\iota_\zeta C_3)+\iota_\zeta G_4~.
\end{alignat}
The first term on the right-hand side pulls back to a total derivative on the fivebrane worldvolume, which 
can be compensated by a gauge transformation. 
The pull-back of the second term on the right-hand side vanishes  
for a bosonic background at $\th=0$, as can be seen by 
(\ref{q}) below and by taking into 
account that the only nonzero components of $G_4$ are $G_{\una\unb\ua\ub}$ 
and $G_{\una\unb\unc\und}$. Similarly, the WZ term transforms under (\ref{a}) as
\bead
\int_{W_6} {\cal L}_{\zeta}(C_6+\frac{1}{2} F_3\wedge C_3)=
\int_{W_6}  {\iota}_{\zeta}(G_7+\frac{1}{2} H_3\wedge G_4)~,
\end{alignat}
where we have dropped a total derivative from the 
integrand. Again, this vanishes 
for a bosonic background at $\th=0$. Finally, the Green-Schwarz 
metric is manifestly invariant under (\ref{a}, \ref{ui}).

Condition (\ref{ui}) can be solved  for $\zeta$, order by order in a $\theta$-expansion. 
By taking the torsion constraints into account, it can be shown that 
\bead
\zeta^{\ua}&=\eta^{\ua}(X)+{\cal O}(\th^2)\nn\\
\zeta^{\una}&=i(\eta\C^{\una}\th)+{\cal O}(\th^3)~,
\label{q}
\end{alignat}
where $\eta^{\ua}$ is a Killing spinor,
\bead
\cD_{\unm}\eta^{\ua}(X)=0~.
\label{m}
\end{alignat}
Transformation (\ref{a}) corresponds to a zero mode  
iff it can be compensated by a $\kappa$-transformation, i.e. iff there 
exists $\kappa^{\ua}(\sigma)$ such that
\bead
\eta^{\ua}(X(\sigma))+\kappa^{\ua}(\sigma)=0~.
\label{p}
\end{alignat}
On the other hand $\kappa$ satisfies 
$\kappa^{\ub}\bar{\C}_{\ub}{}^{\ua}=\kappa^{\ua}$, 
where
\bead
\bar{\C}(\sigma):= \frac{1}{\sqrt{det(\d_r{}^s+i\widetilde{H}_r{}^s)}}
\Big\{
\frac{1}{6!}\frac{\e^{m_1\dots m_6}}{\sqrt{-g}}\C_{m_1\dots m_6}
&+\frac{i}{2}\C_{mnp}\widetilde{H}^{mn}v^p\nn\\
&-\frac{1}{16}\frac{\e^{m_1\dots m_6}}{\sqrt{-g}}
\widetilde{H}_{m_1m_2}\widetilde{H}_{m_3m_4}
\C_{m_5m_6}
\Big\}~,
\label{kproj}
\end{alignat}
so that $\bar{\C}^2=1$. 
Hence (\ref{p}) is equivalent to
\bead
\eta^{\ub}(X(\sigma))(1-\bar{\C}(\sigma))_{\ub}{}^{\ua}=0~,
\label{k}
\end{alignat}
with $\bar{\C}(\sigma)$ evaluated for the bosonic fivebrane superembedding  
in the bosonic background.  
To summarize: the `global' zero modes are given by  
\bead
\th^{\ua}(\sigma)=\eta^{\ua}(X(\sigma))~,
\end{alignat}
where $\eta$ satisfies (\ref{m}), (\ref{k}). Consequently, $\th^{\ua}$ is 
annihilated by $\cD_m=\partial_mX^{\unm}\cD_{\unm}$ and hence obeys the 
Dirac equation on the fivebrane: 
\beal
\C^m\cD_m\th=0~, 
\label{dirac}
\end{align}
which follows from the quadratic part of the fivebrane action 
(\ref{ityu}). 
I.e. `global' 
zero modes give rise to zero modes on the fivebrane. The converse is not generally true.


\subsection*{Supersymmetric cycles in the case of normal  flux}

For a large six-cycle ${\Sigma}$, $X$ can  be approximated by the total space 
of the normal bundle of ${\Sigma}$ in $X$ as in \cite{w}.
 Equivalently, ${\Sigma}$ can be specified 
by a complex vector field $K$ on $X$ such that 
\beal
ds^2(X)
=G_{mn}d\sigma^{m}\otimes d\sigma^{n}+K\otimes K^*   ~,
\label{see}
\end{align}
where $G_{mn}(\sigma)$ is  the metric of 
${\Sigma}$,  and $K^mG_{mn}=0$. We shall normalize 
$K$ as in appendix \ref{sus}, $|K|^2=2$,  
in which case the determinants of the metrics on $X$, ${\Sigma}$ 
are equal.

The kappa-symmetry 
projector simplifies considerably in the case of normal flux. Passing 
to the static gauge and Wick-rotating, condition (\ref{k}) 
can be seen to be equivalent to 
\beal
\Big(1-\frac{ K^mK^{*n}\e_{mn}{}^{m_1\dots m_6}}{2\times 6!\sqrt{G}}\C_{m_1\dots m_6}\Big)\xi=0
~.
\label{iy}
\end{align}
Furthermore, 
using the formul{\ae} in the appendix, 
equation (\ref{iy}) can be rewritten as
\beal
P^+\xi=\xi; ~~~~~ 
P^+:=\frac{1}{2}\Big(1+\frac{1}{2}
K^mK^{*n}\Gamma_{mn}\Gamma_9
\Big)
~.
\label{iyu}
\end{align}
The normal vector 
$K$ is not  a priori holomorphic with respect to the 
complex structure of $X$. However, it is straightforward to 
see from (\ref{iyu}) that 
\beal
J_m{}^nK_n=-iK_m~.
\end{align}
It follows that in the case of normal flux, supersymmetric cycles are antiholomorphic 
cycles.


\subsection{Zero modes}
\label{zeromodes}

We are now ready to come to the analysis of the fermionic zeromodes 
on the worldvolume of the fivebrane. The main result of this 
section is given in (\ref{zmeqs}) below. In the process we make contact 
with the earlier results of \cite{saul, kall}. The form of the Dirac operator in the 
linear approximation was derived in \cite{dira}.

A note on notation: in the remainder of the paper, lower-case Latin letters  
from the middle of the alphabet ($m,n,\dots$) denote 
indices along $X$ (as opposed to indices along 
the fivebrane worldvolume).

\subsection*{Spinors-forms correspondence on $X$}

Using formul{\ae} (\ref{fierzsu}) in appendix \ref{sus} 
we can see that any chiral spinor $\lambda_+$ on $X$ can be expanded as
\beal
\lambda_+=\Phi^{(0,0)}\xi+\Phi^{(2,0)}_{mn}\g^{mn}\xi+\Phi^{(4,0)}_{mnpq}\g^{mnpq}\xi ~,
\end{align}
where $\Phi^{(p,0)}$ is a $(p,0)$-form with respect to the 
complex structure $J$. I.e.  $\Phi^{(2,0)}$ is 
in the $\bf{6}$ of $SU(4)$ and  $\Phi^{(4,0)}$  
is a singlet. 
Similarly in the case of an antichiral spinor $\lambda_-$ we can expand 
\beal
\lambda_-=\Phi^{(1,0)}_{m}\g^{m}\xi+\Phi^{(3,0)}_{mnp}\g^{mnp}\xi ~,
\end{align}
where $\Phi^{(1,0)}$ is 
in the $\bf{4}$ of $SU(4)$ and  $\Phi^{(3,0)}$ 
is in the $\bar{\bf{4}}$. More succinctly, the equations above 
are nothing but the equivalence 
\beal
S_+&\cong \Lambda^{({\rm even}, 0)}\nn\\
S_-&\cong \Lambda^{({\rm odd}, 0)}~,
\end{align}
which can be shown to hold in the case of a Calabi-Yau manifold.

\subsection*{Spinors-forms correspondence on the fivebrane}

We will now assume that the fivebrane wraps a supersymmetric cycle, as 
described above. 
Ignoring the three flat directions for simplicity, 
after gauge-fixing the kappa-symmetry 
the fermions on the worldvolume of the fivebrane transform as sections 
of the tensor product
\beal
S_+\otimes (S_+(N)\oplus S_-(N)) &\cong
\Lambda^{(0,0)}\oplus\Lambda^{(2,0)}\oplus K \oplus(K\otimes\Lambda^{(2,0)}) \nn\\
&\cong \Lambda^{(0,0)}\oplus \Lambda^{(2,0)} \oplus\Lambda^{(0,1)}\oplus\Lambda^{(0,3)}~,
\label{kloi}
\end{align}
where $S_{\pm}(N)$ are the positive-, negative-chirality spin bundles associated 
to the normal bundle $N$ of ${\Sigma}$ in $X$, 
$\Lambda^{(p,0)}$ is the bundle of  
$(p,0)$-forms on ${\Sigma}$, 
and $K$ is the canonical bundle of ${\Sigma}$. The first equivalence above 
can be shown by taking the adjunction formula into account, and the triviality 
of the canonical bundle of $X$. The second equivalence is proven by 
noting that $K\otimes\Lambda^{(3-p,0)}\cong \Lambda^{(0,p)}$, 
as can be seen by contracting with the 
antiholomorphic $(0,4)$-form on $X$.

More explicitly, after gauge-fixing the kappa-symmetry, the physical 
fermion $\theta$ on the world-volume ${\Sigma}$ can be expanded as
\beal
\theta=\epsilon\otimes 
P^+\sum_{p=0}^{4}\Phi^{(p,0)}_{i_i\dots i_p}\g^{{i_i\dots i_p}}\xi~,
\label{ty}
\end{align}
where $\Phi^{(p,0)}\in \Lambda^{(p,0)}$ and $\epsilon$ is a two-component 
spinor in the noncompact directions. Expanding 
\beal
\Phi^{(p,0)}=\widehat{\Phi}^{(p,0)}+\frac{1}{p}K^*\wedge\widehat{\Psi}^{(p-1,0)}~,
\end{align}
where $\iota_K\widehat{\Phi}$, $\iota_K\widehat{\Psi}=0$, 
and substituting $P^+$, 
(\ref{ty}) reads
\beal
\theta=\epsilon\otimes \Big( \widehat{\Phi}^{(0,0)}+ 
\widehat{\Phi}^{(2,0)}_{ij}\gamma^{ij}
+\widehat{\Phi}^{(1,0)}_i\gamma^i
+\widehat{\Phi}_{ijk}^{(3,0)}\gamma^{ijk}
\Big)\xi
~,
\label{koo}
\end{align}
where we have set
\beal
\widehat{\Phi}^{(1,0)}_i&:=\widehat{\Psi}^{(0,0)}K^*_{i}\nn\\
\widehat{\Phi}_{ijk}^{(3,0)}&:=\widehat{\Psi}_{[ij}^{(2,0)}K^*_{k]}~.
\label{leg}
\end{align}
Equation (\ref{koo}) above is the explicit form of (\ref{kloi}).

\subsection*{Zero modes}

The zero modes on the fivebrane satisfy the Dirac equation (\ref{dirac})  
where, after gauge-fixing  
$\theta$ has positive chirality along the fivebrane world-volume, $\theta=P^+\theta$. 
Having explained the spinor-form correspondence, we would now like to rewrite  
the Dirac equation in terms of forms on the fivebrane. First, it would be useful to note the 
following relations:
\beal
(\Pi^{\parallel})_m^rF_{rnpq}\g^m\g^{npq}\theta_-&=0\nn\\
(\Pi^{\parallel})_m^rF_{rnpq}\g^m\g^{npq}\theta_+&=\frac{3}{4}F_{mnpq}\g^{mnpq}\theta_+~,
\label{topi}
\end{align}
where $\theta_{\pm}$ denotes the chirality of $\theta$ along the normal directions, 
and  $\Pi^{\parallel}$ is the projector onto the fivebrane worldvolume defined 
in appendix \ref{kixlh}. 
Since $\theta$ has positive chirality along the fivebrane world-volume, we have 
$\theta_{\pm}=\frac{1}{2}(1\pm\C_9)\theta$. 
It further follows that
\beal
\cD_m\th^{(p,0)}=\epsilon\otimes
\left\{
\begin{array}{ll}
\nabla_m\widehat{\Phi}\xi~,   & ~~~~~p=0\\
\nabla_m\widehat{\Phi}_r\ga^r\xi-\frac{1}{4}\widehat{\Phi}^rF_{rstm}\ga^{st}\xi~,   & ~~~~~p=1\\
\nabla_m\widehat{\Phi}_{rs}\ga^{rs}\xi-\frac{1}{6}\widehat{\Phi}^{rn}F_{rstm}\ga^{st}{}_n\xi~,   & ~~~~~p=2\\
\nabla_m\widehat{\Phi}_{rst}\ga^{rst}\xi-\frac{3}{4}\widehat{\Phi}^{rnp}F_{rstm}\ga^{st}{}_{np}\xi~,   & ~~~~~p=3
\end{array}\right.
~,
\label{koptz}
\end{align}
where we have denoted $\th^{(p,0)}:=\epsilon\otimes
\widehat{\Phi}^{(p,0)}_ {i_1\dots i_p}\ga^{i_1\dots i_p}\xi$. 
Plugging (\ref{koptz})   
into (\ref{dirac}), we obtain
\begin{center}
\fbox{\parbox{11cm}{
%
%
\beal
0&=\Big\{(\nabla^{\parallel})_m\widehat{\Phi}
+4(\nabla^{\parallel})^{ p}\widehat{\Phi}_{pm}\Big\}\g^m\xi\nn\\
0&=\Big\{(\nabla^{\parallel})_{m}\widehat{\Phi}_{n}+6(\nabla^{\parallel})^{ p}\widehat{\Phi}_{pmn}
-\frac{1}{2}F_{mn}{}^{pq}\widehat{\Phi}_{pq}\Big\}\Omega^{mnrs}\g_{rs}\xi^*\nn\\
0&=\Big\{(\nabla^{\parallel})_{m}\widehat{\Phi}_{np}\Big\}\Omega^{mnpq}\g_q\xi^*\nn\\
0&=\Big\{(\nabla^{\parallel})_{m}\widehat{\Phi}_{npq}\Big\}\Omega^{mnpq}
~,\nn
\end{align}
%
%
%
}}
\end{center}
\beal\label{zmeqs}\end{align}
where $(\nabla^{\parallel})_m:= (\Pi^{\parallel})_m^n\nabla_n$, is the covariant derivative 
projected along the fivebrane.  
Passing to complex coordinates, the above can be seen to be equivalent to  
equations (3.6-3.9) of \cite{saul}, or (3.10-3.13) of \cite{kall}.

Following the analysis of \cite{kall}, the space of solutions to 
the above system of equations is spanned by harmonic forms\footnote{\label{foot} 
The forms $\widehat{\Phi}_{I_p}^{(p,0)}$, $p=1,3$, have a leg in the normal bundle, 
see definition (\ref{leg}). More precisely: they are in $H^{0}(\Sigma, K\otimes\Omega^{3-p})$, $p=1,3$. 
Out of these, we can construct 
harmonic forms in $H^{0,p}(\Sigma)\cong H^p(\Sigma, {\cal O})$, by contracting with the antiholomorphic 
fourform on $X$. 
This is just the statement of Serre duality. 
} 
$\{\widehat{\Phi}_{I_{p}}^{(p,0)}; ~p=0\dots 3\}$, 
where in addition the $\widehat{\Phi}^{(2,0)}$s satisfy the constraint
\beal
\mathcal{H}\Big\{F_{mnpq}\widehat{\Phi}^{np}(\Pi^{\parallel})_r^qdx^r\Big\}=0
\label{hcon}
\end{align}
and we have denoted by $\mathcal{H}$ the projector onto the space of harmonic forms. 
The corresponding 
fermion zero modes are of the form 
\beal
\theta= \sum_{p=0}^3\sum_{I_p}
\epsilon^{I_{p}}\otimes X_{I_p}\xi ~,
\label{zm}
\end{align}
where (no summation over $p$)
\beal
X_{I_p}=\left\{  \begin{array}{ll}
\widehat{\Phi}_{I_{p}}^{(p,0)}\gamma_{(p)}, & ~~~p\neq2\\
\widehat{\Phi}_{I_{2}}^{(2,0)}\gamma_{(2)}+\delta\widehat{\Phi}^{(1,0)}_{I_{2}}\gamma_{(1)} 
+\delta\widehat{\Phi}^{(3,0)}_{I_{2}}\gamma_{(3)}  , & ~~~p=2
\end{array}\right. 
;~~~
I_p=\left\{  \begin{array}{ll}
1,\dots, h^{p,0}({\Sigma}), & ~~~p\neq2\\
1,\dots,n, & ~~~p=2
\end{array}\right.~,
\end{align}
the $\widehat{\Phi}_{I_{p}}^{(p,0)}$s are harmonic 
and $\{\delta\widehat{\Phi}^{(1,0)}_{I_{2}}$, $\delta\widehat{\Phi}^{(3,0)}_{I_{2}}\}$ is 
a special solution of the inhomogeneous equation
\beal
(\nabla^{\parallel})^+_{[m}\widehat{\Phi}_{n]}+6(\nabla^{\parallel})^{ p}\widehat{\Phi}_{pmn}
=\frac{1}{2}F_{mn}{}^{pq}\widehat{\Phi}_{I_2,pq}
~.
\label{opuio}
\end{align}
In the above, $n$ is the number of harmonic (2,0) forms on ${\Sigma}$ 
which in addition satisfy the constraint (\ref{hcon}); 
the $\epsilon^{I_{p}}$s are spinors in the  $\bf{2}$ of $Spin(3)$ (after 
Wick-rotating to Euclidean signature). 
Note that (\ref{opuio}) implies condition (\ref{hcon}). 
The authors of \cite{kall} 
define a flux-dependent generalization of the arithmetic genus:
\beal
\chi_F:=h^{0,0}-h^{1,0}+n-h^{3,0}~.
\label{ketal}
\end{align}
%


\section{Instanton contributions}
\label{instantoncontributions}

We can now proceed to the computation of the instanton contributions to the coupling 
(\ref{fbils}). The main result of the paper is arrived at in this section: 
instantons with four fermionic zeromodes do not contribute to the superpotential.


\subsection{Gravitino Kaluza-Klein reduction}
\label{gravitinokkreduction}

Before proceeding to integrate over the fermion zeromodes, 
we will need the Kaluza-Klein ansatz for the gravitino entering the 
vertex operator $V$ in (\ref{grv}). 
As already discussed in the introduction, only terms 
which depend on the descendants of the linear multiplets contribute to the superpotential. Hence, the 
relevant part  of the Kaluza-Klein ansatz for the gravitino reads
\beal
\left\{
\begin{array}{l}
\Psi_{\mu}=i(\omega_I\cdot J)~\gamma_{\mu}\chi^I\otimes\xi^* +{\rm c.c.}  \\
\Psi_m = \chi^I\otimes
\omega_{I,mp}\ga^p\xi^*+{\rm c.c.}~; ~~~~~I=1,\dots b_2~,
\end{array}\right.
\label{kkgr}
\end{align}
where 
the $\chi^{I}$s are complex spinors in the  $\bf{2}$ of $Spin(3)$, 
and $\omega_I\in H^{2}(X,\mathbb{R})$. 
As is straightforward 
to see, the eleven-dimensional gravitino equation, $\Gamma^M\cD_{[M}\Psi_{N]}=0$, 
is satisfied if $\chi^{I}$ is a massless three-dimensional fermion,
\beal
\slsh\nabla\chi^I=0~,
\end{align}
provided 
\beal
\omega_I\lrcorner F=0~.
\label{34}
\end{align}
The implications of this condition were discussed extensively in the introduction. 
In this picture, $\chi^{I}$ is massless if it corresponds to a zero eigenvalue 
of the matrix $T_{IJ}$ (in a diagonal basis). Alternatively this can be seen as follows. 
%
The quadratic part of the three-dimensional action for the $\chi^I$s 
comes from the dimensional reduction of the quadratic-gravitino term 
in the eleven-dimensional supergravity action
\beal
\int{
d^{11}x\sqrt{g_{11}} \Psi_M\C^{MNP}\cD_N\Psi_P
}~.
\end{align}
Plugging the Kaluza-Klein ansatz (\ref{kkgr}) in the action above, we obtain
\beal
{\rm Vol}(X)\int{
d^{3}x\sqrt{g_{3}} \Big(
D_{IJ}\bar{\chi}^I\slsh\nabla\chi^J-\frac{4}{9}T_{IJ}\bar{\chi}^I\chi^J
\Big)}~,
\label{3daction}
\end{align}
where
\beal
D_{IJ}&:= \int_X \Big(
\omega_I\wedge\star\omega_J+\frac{2}{3}~\omega_I\wedge\omega_J\wedge J\wedge J
\Big) 
~
\label{irw}
\end{align}
and the Hodge star is with respect to the 
metric of the Calabi-Yau fourfold. In the above we have made use of the identity
\beal
\star(\omega_I\wedge\omega_J\wedge J\wedge J)
=\frac{1}{2}\Big\{(\omega_I\cdot J)(\omega_J\cdot J)-2(\omega_I\cdot\omega_J)\Big\}~,
\end{align}
which can be proven with the help of  
(\ref{jids}). As advertised, 
massless fermions correspond to zero eigenvalues of $T_{IJ}$. 


We remark that in (\ref{3daction}) there is no coupling of the form
\beal
{\rm Vol}(X)\int
d^{3}x\sqrt{g_{3}}\Big( 
W_{IJ}\chi^I\chi^J
+{\rm c.c.}
\Big)~.
\label{poten}
\end{align}
In the following we will investigate whether such a term is 
generated by instanton contributions. In the context 
of three-dimensional supersymmetric field theory the fact that such 
a term can indeed be generated by instanton effects, was demonstrated in 
\cite{wittenold}.

\subsection{Two zeromodes}
\label{ofrzm}

Before coming to the subject of instantons with four fermionic zeromodes in the next subsection, 
we will briefly comment on the case of instantons with two zeromodes (corresponding to the fivebrane wrapping rigid, 
isolated cycles). As can be seen from (\ref{zm}), 
there are  always two zero modes corresponding to $p=0$: 
\beal
\theta= \epsilon\otimes\xi~.
\label{oipi}
\end{align}
These are the 
zero modes which come from the supersymmetry of the Calabi-Yau background\footnote{
In three-dimensional nomenclature the supersymmetry of the 
background is ${\cal N}=2$  
(equivalently: ${\cal N}=1$ in four dimensions), i.e. 
four real supercharges. The instanton breaks half the 
supersymmetries, as can be seen from (\ref{iyu}). 
Note that $\xi$ in (\ref{oipi}) is complex 
and $\epsilon$ is a 
 spinor in the  $\bf{2}$ of $Spin(3)$. Henceforth 
we are complexifying our notation for $\theta$, $\Psi_m$, $V$. 
At any rate, $\theta$ must be complexified in order to pass to 
Euclidean signature.}.  
We would like to compute the instanton contribution of these zeromodes to the superpotential. 
First, we need to  define the integration over fermion zeromodes:
\beal
\int d^2\epsilon~\epsilon^{\alpha}\epsilon^{\beta}:=C^{\alpha\beta}~,
\label{zint}
\end{align}
%
%
where $C$ in the equation above is the charge-conjugation matrix in three dimensions. 
It follows that 
\beal
\int d^2\epsilon~(\chi\epsilon)(\epsilon\psi)=(\chi\psi)~,
\end{align}
for any two three-dimensional spinors $\chi$, $\psi$ 
in the  $\bf{2}$ of $Spin(3)$. To simplify the presentation, 
we are using the notation $(\chi\psi):=(\chi^{Tr}C\psi)$.

Integrating over the zeromodes using the above prescription, we find that the instanton induces a 
two-fermion coupling of the form
\beal
\chi^I\chi^J\int [DZ'(\sigma)]v_Iv_{J} 
 e^{-S_{PST}[Z(\sigma); g,C,\Psi]}+{\rm c.c.} ~,
\label{2zm}
\end{align}
where
\beal
v_I&:=2i\int_{{\Sigma}}J\wedge J \wedge\omega_I
\label{vdef}
\end{align}
and the path integration above does not include the zeromodes. 
In (\ref{vdef}) all the forms should be understood as pulled-back to  ${\Sigma}$. 
In particular the pull-back of the 
almost complex structure to ${\Sigma}$ can be identified with 
$\widehat{J}$, which is discussed from the point of view of the 
induced $SU(3)$ structure on ${\Sigma}$ in appendix \ref{kixlh}. Note that in the formula above 
the primitive part of $\omega_I$ is projected out. 

We are not going to elaborate on the one-loop determinants, as this lies outside the 
main focus of this paper. 
The result of the 
integration over the bosonic coordinates should be obtainable using 
techniques similar to \cite{hm}. The 
integration over the fermionic variables is proportional to the 
determinant of the flux-dependent 
Dirac operator $\gamma^m\cD^{\parallel}_m$ (away from its kernel), 
as follows from equation (\ref{ityu}).

\subsection{Four zeromodes}
\label{frzm}

In the presence of four zeromodes there  are the following possibilities 
which we will examine in turn: either $h^{0,0}=n=1$ (corresponding 
to $\chi_F=2$) or 
$h^{0,0}= h^{p,0}=1$, where $p$ is odd (corresponding 
to $\chi_F=0$). Recall that $n$ is the number of harmonic (2,0) forms on ${\Sigma}$ 
which in addition satisfy the constraint (\ref{hcon}). As we will see, no superpotential 
is generated in either case. Since $\chi_F\neq 1$ in all cases, we conclude that our 
result does not rule out the possibility that in the presence of flux the arithmetic genus criterion 
should be replaced by the condition $\chi_F=1$. 

$~\bullet h^{0,0}=n=1$

In this case we have $\chi_F=2$. Let us substitute 
 the Kaluza-Klein ansatz (\ref{kkgr}) and the expression for the zeromodes, 
\beal
\theta=
\epsilon\otimes\xi+ \zeta\otimes\Big(
\widehat{\Phi}_{mn} \g^{mn}+\delta\widehat{\Phi}_{m}\g^{m}
+\delta\widehat{\Phi}_{mnp}\g^{mnp}
\Big)\xi~, 
\end{align}
into equation (\ref{grv}) for the gravitino vertex operator. 
Integrating over the zeromodes using (\ref{zint}) we get, up to a total 
worldvolume derivative,
\beal
\int d^2\epsilon ~d^2\zeta~V V =
\chi^I\chi^J v_I w_{J}~,
\label{4zm}
\end{align}
where $v_I$ was defined in (\ref{vdef}) above and 
\beal
w_{J}&:=
\frac{2}{9}\int_{{\Sigma}}
\widehat{\Theta}\wedge\widehat{\Phi}\wedge\omega_J
~.
\label{ji}
\end{align}
%
The object $\widehat{\Theta}$ is defined by
\beal
\widehat{\Theta}_{mn}:=\Omega_{mnpq}F^{pq}{}_{rs}\widehat{\Phi}^{rs}
\label{thdefi}
\end{align}
and is a (0,2)-form on ${\Sigma}$. (Recall that in our conventions 
$\Omega$ is antiholomorphic). 
In deriving this result, we had to perform some  tedious but 
straightforward gamma-matrix algebra making repeated use of the formul\ae {} in the 
appendices \ref{gammapp}, \ref{sus}, especially 
equations (\ref{bfive}, \ref{usefids}). 
Moreover we have taken into account the normal flux condition 
%
%
and we have implemented (\ref{dkn}),  as discussed in the introduction.
%
%
%
%
%
%
%

In the following we show that the right-hand side of (\ref{ji}) vanishes; 
no instanton-induced superpotential is generated in this case. Before demonstrating this 
fact however, let us note that the following group-theoretical reasoning can be used to gain 
insight into the result (\ref{4zm}). As follows from the 
form of the vertex operator, the integration over the 
zeromodes receives three kinds of contributions:
\beal
\chi^I\chi^J v_I\otimes \omega_J\otimes F\otimes
(\widehat{\Phi}^{(2,0)}+ \delta\widehat{\Phi}^{(1,0)}+ \delta\widehat{\Phi}^{(3,0)} )^{2\otimes_s}
~,
\label{a111}
\end{align}
coming from terms of the form $VV\propto (\Psi_m\C^m\th)(\Psi\th^3)F$,
\beal
\chi^I\chi^J
v_I\otimes \nabla\omega_J\otimes(\widehat{\Phi}^{(2,0)}+ 
\delta\widehat{\Phi}^{(1,0)}+ \delta\widehat{\Phi}^{(3,0)} )^{2\otimes_s}~,
\label{a211}
\end{align}
coming from terms of the form $VV\propto (\Psi_m\C^m\th)(\nabla\Psi\th^3)$, 
and
\beal
\chi^I\chi^Jv_I\otimes \omega_J\otimes
(\widehat{\Phi}^{(2,0)}+ \delta\widehat{\Phi}^{(1,0)}+ \delta\widehat{\Phi}^{(3,0)} )
\otimes
\nabla
(\widehat{\Phi}^{(2,0)}+ \delta\widehat{\Phi}^{(1,0)}+ \delta\widehat{\Phi}^{(3,0)} )
\label{a311}~,
\end{align}
coming from terms of the form $VV\propto (\Psi_m\C^m\th)(\Psi\th^2\nabla\th)$. 
Contributions of the type (\ref{a111}) transform in the\footnote{
In the following we are using the Dynkin notation for $A_3$.}
$$
\Big((000)\oplus(101)\Big)\otimes (020)\otimes\Big(
(010) \oplus(100)\oplus(001) \Big)^{2\otimes_s}
$$
of $SU(4)$. There are exactly three scalars in the decomposition of the 
tensor product above. These  we can write explicitly as:
\beal
S_1&:=\chi^I\chi^Jv_I\omega_{J, mn}\Omega^{mpij}
F_{ijqr}\widehat{\Phi}^{qr}\widehat{\Phi}_{p}{}^n\nn\\
S_2&:=\chi^I\chi^Jv_I(\omega_J\cdot J)
\Omega^{mpij}
F_{ijqr}\widehat{\Phi}^{qr}\widehat{\Phi}_{mp}\nn\\
S_3&:=\chi^I\chi^Jv_I\delta\widehat{\Phi}^i\delta\widehat{\Phi}^{jk}{}_m
\Omega_{ijkn}F^{mnpq}\omega_{J, pq}~.
\end{align}
The last one, however, vanishes by virtue of equation (\ref{34}). 
Moreover, using equation (\ref{opuio}), the scalars 
$S_{1,2}$ can be expressed as a linear combination of 
$R_{1}, \dots R_7$ defined in equation (\ref{taro}) below:
\beal
S_1&=-2R_2+4R_5-4R_6\nn\\
S_2&=2R_4-8R_7
~.
\end{align}
In deriving the above we have used the identity
\beal
\delta\widehat{\Phi}_{qrs}\Omega^{rsmp}=-\frac{2}{3}\Omega^{ijk[m}(\Pi^+)_q{}^{p]}
\delta\widehat{\Phi}_{ijk}
~,
\end{align}
which can be proved using (\ref{bfive}). A direct computation 
of the terms of the form (\ref{a111}), yields the contribution 
\beal
\frac{2i}{9}S_1-\frac{1}{18}S_2=-\frac{4i}{9}(R_2-2R_5+2R_6)- \frac{1}{9}(R_4-4R_7)
\label{contr1}
\end{align}
to the zeromode integral (\ref{4zm}). The linear combination above can 
be written in a more elegant way by noting that 
\beal
iS_1-\frac{1}{4}S_2=\chi^I\chi^Jv_I
\star (\widehat{\Theta}\wedge\widehat{\Phi}\wedge\omega_J)~,
\label{525}
\end{align}
where the Hodge star is along ${\Sigma}$. In proving (\ref{525}) we have 
made use of equation (\ref{b0}).

Taking into account that $\omega_I$ is a harmonic (1,1) form 
and that therefore $(\omega_I\cdot J)$ is a constant\footnote{\label{ext} 
A direct computation reveals that it is in fact  
$\widehat{\omega}^I$ rather than 
$\omega_I$, where the hat denotes 
the pull-back to $\Sigma$, which appears in the various invariants 
of this section. However, 
%
using the inclusion map
$$
\iota^*:~H^{p,q}(X,\mathbb{R})\longrightarrow H^{p,q}({\Sigma},\mathbb{R})~,
$$
%
we can  think of ${\omega}^I$ as the extension to $X$  
of the  harmonic form $\widehat{\omega}^I$ on $\Sigma$ \cite{gh}. 
In the text, we do not make an explicit distinction 
between  $\omega_I$ and $\widehat{\omega}^I$. See also the next footnote. 
}, 
it follows that 
$\nabla\omega_I$ transforms in the $(201)\oplus(102)$ of $SU(4)$. Hence, 
contributions of the type (\ref{a211}) transform in the 
$$
\Big((201)\oplus(102)\Big)\otimes \Big(
(010) \oplus(100)\oplus(001) \Big)^{2\otimes_s}
$$
of $SU(4)$. As there are no scalars in the decomposition of the 
tensor product above, we conclude that these terms vanish.

Taking into account that $\widehat{\Phi}^{(2,0)}$ is a harmonic 
(2,0) form on a K\"{a}hler manifold, it follows 
that $\nabla\widehat{\Phi}^{(2,0)}$ transforms in the $(110)$ of $SU(4)$. 
Similarly, $\nabla\delta\widehat{\Phi}^{(1,0)}$ 
transforms in the 
$(000)\oplus(200)\oplus(010)\oplus(101)$ of $SU(4)$. 
Finally, taking into account the last of equations (\ref{zmeqs}), it 
follows that $\nabla\delta\widehat{\Phi}^{(3,0)}$ transforms in the 
$(010)\oplus(101)\oplus(002)$ of $SU(4)$. 
Putting everything together, it follows that 
contributions of the type (\ref{a311}) transform in the 
\beal
\Big((000)\oplus(101)\Big)\otimes \Big((010)\oplus&(100)\oplus(001)   \Big)\nn\\
\otimes &\Big(
(110)\oplus(000)\oplus(200)
\oplus 2(010)\oplus 2(101)\oplus(002)
\Big)\nn
\end{align}
of $SU(4)$. There are exactly seven scalars in the decomposition of the 
tensor product above: 
one coming from $\nabla\widehat{\Phi}^{(2,0)}$, 
three from $\nabla\delta\widehat{\Phi}^{(1,0)}$ 
and three from $\nabla\delta\widehat{\Phi}^{(3,0)}$. 
These can be written explicitly as
\beal
R_1&:=\chi^I\chi^Jv_I\nabla^m\widehat{\Phi}_{ij}\Omega^{ijpq}\delta\widehat{\Phi}_{p}
\omega_{J,qm}\nn\\
R_2&:=\chi^I\chi^Jv_I\nabla_m\delta\widehat{\Phi}_n\Omega^{mnij}\omega_{J,ip}
\widehat{\Phi}^{p}{}_{j}\nn\\
R_3&:=\chi^I\chi^Jv_I\nabla^m\delta\widehat{\Phi}_n
\Omega^{nijk}\omega_{J,km}\widehat{\Phi}_{ij}\nn\\
R_4&:=\chi^I\chi^Jv_I(\omega_J\cdot J)\nabla_m\delta\widehat{\Phi}_n
\Omega^{mnij}\widehat{\Phi}_{ij}\nn\\
R_5&:=\chi^I\chi^Jv_I\nabla^m\delta\widehat{\Phi}_{ijk}\Omega^{ijkq}
\omega_{J,qp}\widehat{\Phi}^{p}{}_{m}\nn\\
R_6&:=\chi^I\chi^Jv_I\nabla^m\delta\widehat{\Phi}_{ijk}\Omega^{ijkq}
\omega_{J,mp}\widehat{\Phi}^{p}{}_{q}\nn\\
R_7&:=\chi^I\chi^Jv_I(\omega_J\cdot J)
\nabla^m\delta\widehat{\Phi}_{ijk}\Omega^{ijkq}
\widehat{\Phi}_{qm}
~.
\label{taro}
\end{align}
A direct computation of the terms of the form (\ref{a311}), yields the contribution 
\beal
-4i(R_1+R_3+2R_5)
\label{contr2}
\end{align}
to the zeromode integral (\ref{4zm}).

Putting the contributions (\ref{contr1}, \ref{contr2}) together, 
we arrive at equation (\ref{4zm}). Note that the 
invariants $R_4, \dots R_7$ as well as the linear combinations 
$R_1+2R_2$ and $R_1+R_3$, can be written as total derivatives. This can 
readily be seen by 
taking into account that $\Omega$ is covariantly constant while $\omega$, $\widehat{\Phi}$ 
are harmonic\footnote{
Note that in general the pull-back of the 
Christoffel connection from the total space $X$ 
to the base $\Sigma$,   
$(\nabla^{\parallel})_m$, {\it cannot}
be identified with the Christoffel connection $\widehat{\nabla}_m$ 
associated with the metric on $\Sigma$. However if $\widehat{S}$ is an  
arbitrary $p$-form on $\Sigma$ whose extension to $X$ is $S$, we have
$$
(\nabla^{\parallel})_m{S}^{mm_2\dots m_p}=
\nabla_m{S}^{mm_2\dots m_p}=
(\widehat{\nabla})_m\widehat{S}^{mm_2\dots m_p}~.
$$
The first equality follows from (\ref{tbb}). The second equality follows  
from $\C_{mn}^n=g^{-1/2}\partial_mg^{1/2}$ 
and the fact that the determinants of the metrics $X$, $\Sigma$ are equal, as can be 
seen from the explicit form of the fibration (\ref{explfibr}).
}.  
It follows that the total contribution can be cast in the form 
$\propto R_2$+total derivative. 
On the other hand, up to a total derivative, $R_2$ is proportional to 
the right-hand-side of (\ref{525}), as follows from 
(\ref{contr1},\ref{525}). 


We are now ready to  show that the left-hand-side of (\ref{ji}) vanishes identically. 
First note that, as follows from (\ref{hcon}) or (\ref{opuio}), 
the projection of $\widehat{\Theta}$ onto the space of harmonic forms on 
${\Sigma}$ vanishes: ${\cal H}\{\widehat{\Theta}\}=0$. It follows that
\beal
\int_{{\Sigma}}\widehat{\Theta}\wedge\widehat{\Phi}\wedge J=0~, 
\label{gnv}
\end{align}
since $\widehat{\Phi}\wedge J$ is harmonic (this can 
be seen by noting that $\star\widehat{\Phi}=\widehat{\Phi}\wedge J$). 
Varying this equation with respect to the K\"{a}hler structure, 
$\phi^I\rightarrow \phi^I+\delta \phi^I$, we get
\beal
\int_{\Sigma}\frac{\delta\widehat{\Theta}}{\delta\phi^I}\wedge\widehat{\Phi}\wedge J
+ \int_{{\Sigma}}\widehat{\Theta}\wedge\widehat{\Phi}\wedge \omega_I=0~.
\label{vrtrr}
\end{align}
Furthermore, under a K\"{a}hler-structure variation the metric transforms as 
\beal
\delta g_{mn}&=\sum_I \delta\phi^I\omega_{I,mp}J_n{}^p~.
\end{align}
Note that the right-hand side above is automatically symmetric in the indices $m$, $n$. 
Taking the above into account together with the fact that 
$S_2$ is a total worldvolume derivative 
it follows that
\beal
\int_{\Sigma}\frac{\delta\widehat{\Theta}}{\delta\phi^I}\wedge\widehat{\Phi}\wedge J=0~.
\label{asd}
\end{align}
In the derivation we made use of the identity
\beal
\widehat{\Phi}^{mn}\Omega_{mnpq}\widehat{\Phi}^{rs}F_{rs}{}^{qt}\omega_{I,pt}=-
\widehat{\Phi}^{mn}\Omega_{mn}{}^{pq}\widehat{\Phi}_s{}^{t}F_{pq}{}^{sr}\omega_{I,rt}
~.
\end{align}
From (\ref{vrtrr}, \ref{asd}) it finally follows that the right-hand side of (\ref{ji}) vanishes, as 
advertised.

No potential is generated in the remaining cases  either, as we now show.

$~\bullet h^{0,0}=h^{1,0}=1$

In this case we have $\chi_F=0$. 
As can be verified by direct computation, 
no potential is generated in this case. 
The easiest way to arrive at this result is by 
the following group-theoretical argument. It follows from the 
form of the vertex operator that the integration over the 
zeromodes receives three kinds of contributions:
\beal
\chi^I\chi^Jv_I\otimes \omega_J\otimes F\otimes\widehat{\Phi}^{(1,0)}\otimes\widehat{\Phi}^{(1,0)}~,
\label{c111}
\end{align}
coming from terms of the form $VV\propto (\Psi_m\C^m\th)(\Psi\th^3)F$,
\beal
\chi^I\chi^Jv_I\otimes \nabla\omega_J\otimes\widehat{\Phi}^{(1,0)}\otimes\widehat{\Phi}^{(1,0)}~,
\label{c211}
\end{align}
coming from terms of the form $VV\propto (\Psi_m\C^m\th)(\nabla\Psi\th^3)$, and 
\beal
\chi^I\chi^Jv_I\otimes \omega_J\otimes\widehat{\Phi}^{(1,0)}\otimes
\nabla\widehat{\Phi}^{(1,0)}
\label{c311}~,
\end{align}
coming from terms of the form $VV\propto (\Psi_m\C^m\th)(\Psi\th^2\nabla\th)$. 
Contributions of the type (\ref{c111}) transform in the 
$$
\Big((000)\oplus(101)\Big)\otimes (020)\otimes (100)^{2\otimes_s}
$$
of $SU(4)$. As there are no scalars in the decomposition of the 
tensor product above, we conclude that these terms vanish.

Taking into account that $\omega_I$ is a harmonic (1,1) form, it follows that 
$\nabla\omega_I$ transforms in the $(201)\oplus(102)$ of $SU(4)$. Hence, 
contributions of the type (\ref{c211}) transform in the 
$$
\Big((201)\oplus(102)\Big)\otimes (100)^{2\otimes_s}
$$
of $SU(4)$. As there are no scalars in the decomposition of the 
tensor product above, we conclude that these terms vanish.

Taking into account that $\widehat{\Phi}^{(1,0)}$ is harmonic, it follows 
that $\widehat{\Phi}^{(1,0)}$ transforms in the $(200)$ of $SU(4)$. Hence, 
contributions of the type (\ref{c311}) transform in the 
$$
\Big((000)\oplus(101)\Big)\otimes (100)\otimes (200)
$$
of $SU(4)$. As there are no scalars in the decomposition of the 
tensor product above, we conclude that these terms vanish.


$\bullet$ $h^{0,0}=h^{3,0}=1$

In this case we have $\chi_F=0$. As in the previous case, 
no potential is generated. This can be shown {\it e.g.} by 
the same type of group-theoretical reasoning as before.

\section{Discussion}

Taking advantage of the recent 
progress in explicit theta-expansions in eleven-dimensional superspace \cite{t}, 
we have performed a computation of the contribution of fivebrane instantons 
with four fermionic zeromodes in M-theory compactifications 
on Calabi-Yau fourfolds with (normal) flux. The calculus of fivebrane 
instantons in M-theory is still largely unexplored, and we hope that our computation 
will initiate a more extensive study of these phenomena directly in M-theory.

We have found that no superpotential is 
generated in this case -- a result which is compatible 
with replacing the arithmetic genus criterion 
by the condition $\chi_F=1$, where $\chi_F$ is the flux-dependent 
`index' of \cite{kall}. It would be interesting to reexamine this statement when the condition 
of normal flux is relaxed.

It would be desirable to explore the obvious generalizations of our computation: 
fivebrane instanton contributions to  non-holomorphic couplings, and/or 
contributions to higher-derivative and multi-fermion couplings as in \cite{bw}. 
The expansions of \cite{t} can also be used to study instantons with more than four zeromodes.

So far the precise 
relation between instanton calculus in M-theory \cite{bbs, hm} and the rules of D-instanton 
computations in string theory put forward in \cite{gg, bill, blum}, has not been clearly spelled out. 
Understanding this 
relation may help clarify some of the conceptual issues associated with the M-theory calculus, see {\it e.g.}  
\cite{hm}. This would be another interesting possibility for future investigation. 

Last but not least, it is important to address the reservations, discussed in the introduction,
 about the fivebrane action of \cite{pst} 
and to incorporate the topological considerations of \cite{beloa, belob} in 
a supersymmetric context\footnote{I would like to thank Greg Moore for correspondence on this point.}. 

\vfill\break

\section*{Acknowledgment}

I am indebted to Greg Moore for encouragement, correspondence and 
valuable comments on several previous versions of 
the manuscript. 
I am also grateful to 
Ralph Blumenhagen, 
Michael Haack, Peter Mayr and Henning Samtleben for useful discussions and correspondence.

\appendix


\section{Gamma-matrix identities}
\label{gammapp}

The gamma matrices in eight dimensions have the following properties

{\it Symmetry}:
\beal
(C\ga_{(n)})^{Tr}=(-)^{\frac{1}{2}n(n-1)}C\ga_{(n)}~,
\end{align}
where $C$ is the charge-conjugation matrix.

{\it Hodge duality}:
\beal
\star\ga_{(n)}=(-)^{\frac{1}{2}n(n+1)}\ga_{(8-n)}\ga_9~,
\end{align}
where $\ga_9$ is the chirality matrix.

{\it Complex conjugation}:
\beal
\ga_{(n)}^*=C\ga_{(n)}C^{-1}~.
\end{align}

{\it Reality}:

A Majorana-Weyl spinor $\xi_{\pm}$ in eight dimensions, where the subscript 
denotes the chirality, satisfies
\beal
\xi_{\pm}^{\dagger}=\pm\xi_{\pm}^{Tr}C~,
\end{align}
which, together with the complex conjugation above, imply
\beal
(\ga_{(n)}\xi_{\pm})^*=\pm C\ga_{(n)}\xi_{\pm}  ~.
\end{align}

{\it Decomposition} $10\rightarrow3 +8$:

We decompose the eleven-dimensional matrices as follows:
\beal
\C^{\mu}&=\ga^\mu\otimes\ga_9~~,~~~~~\mu=0,1,2\nn\\
\C^{m}&={1}_{2}\otimes\ga^m~,~~~~~m=3,\dots 10~.
\end{align}
The eleven-dimensional charge-conjugation matrix decomposes as
\beal
C_{11}=C_3\otimes C_8\ga_9~,
\end{align}
where $\ga_9$ is the chirality matrix in eight dimensions and $C_3$, $C_8$ are 
the  charge-conjugation matrices in three, eight dimensions respectively.

\section{$SU(4)$ structure}
\label{sus}

The existence of a nowhere-vanishing positive-chirality complex spinor $\xi$ on $X$, 
implies the reduction of the structure group to $SU(4)$. The 
$SU(4)$ structure can be equivalently specified in terms of a complex 
self-dual fourform 
$\Omega$ and an almost complex structure $J$ satisfying
\beal
J\wedge\Omega&=0\nn\\
\Omega\wedge\Omega^*&=\frac{2}{3}J^4 ~.
\end{align}
In terms of $\xi$ bilinears we have
\beal
J_{mn}&=i\xi^{\dagger}\ga_{mn}\xi\nn\\
\Omega_{mnpq}&=\xi^{Tr}\ga_{mnpq}\xi
\end{align}
and we have normalized $\xi^\dagger\xi=1$. Using the almost complex structure 
we can define the projectors 
\beal
(\Pi^{\pm})_m{}^n:=\frac{1}{2}(\delta_{m}{}^{n}\mp i J_m{}^n)
\end{align}
with respect to which $\Omega$ is antiholomorphic
\beal
(\Pi^{-})_m{}^i\Omega_{inpq}=\Omega_{mnpq}~; ~~~~~(\Pi^{+})_m{}^i\Omega_{inpq}=0 ~.
\end{align}
The following useful identities can be proved e.g. by Fierzing
\beal
\frac{1}{4!\times 2^4}~&\Omega_{rstu}\Omega^{*rstu}=1\nn\\
\frac{1}{6\times 2^4}~&\Omega_{irst}\Omega^{*mrst}
=(\Pi^-)_{i}{}^{m}\nn\\
\frac{1}{4\times 2^4}~&\Omega_{ijrs}\Omega^{*mnrs}
=(\Pi^-)_{[i}{}^{m}(\Pi^-)_{j]}{}^{n}\nn\\
\frac{1}{6\times 2^4}~&\Omega_{ijkr}\Omega^{*mnpr}
=(\Pi^-)_{[i}{}^{m}(\Pi^-)_{j}{}^{n}(\Pi^-)_{k]}{}^{p}\nn\\
\frac{1}{4!\times 2^4}~&\Omega_{ijkl}\Omega^{*mnpq}
=(\Pi^-)_{[i}{}^{m}(\Pi^-)_{j}{}^{n}(\Pi^-)_{k}{}^{p}(\Pi^-)_{l]}{}^{q}~,
\label{bfive}
\end{align}
and
\beal
\ga_m\xi&=(\Pi^-)_{m}{}^{n}\ga_n\xi\nn\\
\ga_{mn}\xi&=-iJ_{mn}\xi -\frac{1}{8}\Omega_{mnpq}\ga^{pq}\xi^*   \nn\\
\ga_{mnp}\xi&=-3iJ_{[mn}\ga_{p]}\xi 
-\frac{1}{2}\Omega_{mnpq}\ga^q\xi^*\nn\\
\ga_{mnpq}\xi&=-3J_{[mn}J_{pq]}\xi +\frac{3i}{4}J_{[mn}\Omega_{pq]ij}\ga^{ij}\xi^*   
+\Omega_{mnpq}\xi^*
~.
\label{fierzsu}
\end{align}
The action of $\gamma_{m_1\dots m_p}$, $p\geq 5$, on $\xi$ can be related to 
the above formul{\ae}, using 
the Hodge properties of gamma matrices given in appendix \ref{gammapp}. 
From the above it follows that
\beal
\xi^{\dagger}\xi=1; &~~~~~\xi^{Tr}\xi=0\nn\\
\xi^{\dagger}\ga_{mn}\xi=-iJ_{mn}; &~~~~~\xi^{Tr}\ga_{mn}\xi=0\nn\\
 \xi^{\dagger}\ga_{mnpq}\xi=-3J_{[mn}J_{pq]}; &~~~~~\xi^{Tr}\ga_{mnpq}\xi=\Omega_{mnpq}\nn\\
\xi^{\dagger}\ga_{mnpqrs}\xi=15iJ_{[mn}J_{pq}J_{rs]}; &~~~~~\xi^{Tr}\ga_{mnpqrs}\xi=0\nn\\
\xi^{\dagger}\ga_{mnpqrstu}\xi=105J_{[mn}J_{pq}J_{rs}J_{tu]}; &~~~~~\xi^{Tr}\ga_{mnpqrstu}\xi=0 ~,
\label{usefids}
\end{align}
where we have made use of the identities
\beal
\varepsilon_{mnpqrstu}J^{rs}J^{tu}&=24J_{[mn}J_{pq]}\nn\\
\varepsilon_{mnpqrstu}J^{tu}&=30J_{[mn}J_{pq}J_{rs]}\nn\\
\varepsilon_{mnpqrstu}&=105J_{[mn}J_{pq}J_{rs}J_{tu]}
~.
\label{jids}
\end{align}
Note that the bilinears 
$\xi^{Tr}\ga_{(p)}\xi$, ~$\xi^{\dagger}\ga_{(p)}\xi$, vanish for $p$ odd. 
Finally, the last line of equation (\ref{bfive}) together with the last line of the 
equation above imply
\beal
\Omega_{[ijkl}\Omega^*_{mnpq]}=\frac{8}{35}\varepsilon_{ijklmnpq}~.
\end{align}

\subsection{$SU(4)$ vs $SU(3)$}
\label{kixlh}
In the case where there exists a nowhere-vanishing complex vector $K$, 
one can construct a corresponding nowhere-vanishing 
negative-chirality complex spinor 
\beal
\xi_-:=K^m\gamma_m\xi~. 
\end{align}
This implies the reduction of the structure group of $X$ 
to $SU(3)$.  
Without loss of generality 
we can take $K$ to be antiholomorphic with respect to the 
almost complex structure $J$,
\beal
(\Pi^-)_m{}^nK_n=K_m; ~~~~~~ (\Pi^+)_m{}^nK_n=0~,
\end{align}
and to satisfy 
\beal
K_mK^m=0; ~~~~~~ K^*_mK^m=2~.
\end{align}
The $SU(3)$ structure can be given in terms 
of an antiholomorphic (0,3) form $\widehat{\Omega}$ and 
a (1,1) form $\widehat{J}$ defined by
\beal
J&=\widehat{J} -\frac{i}{2}K\wedge K^*\nn\\
\Omega&=iK\wedge\widehat{\Omega}~,
\end{align}
which satisfy 
$\iota_K\widehat{J}, \iota_K\widehat{\Omega}, \iota_{K^*}\widehat{\Omega}=0$. 
Moreover, we can define 
antiholomorphic projectors ($\widehat{\Pi}$) 
with respect to the structure $\widehat{J}$:
\beal
(\Pi^-)_m{}^n=(\widehat{\Pi}^-)_m{}^n+\frac{1}{2}K_mK^{*n}~.
\label{pihat}
\end{align}
The complex vector $K$ specifies an almost product structure given by 
\beal
R_m{}^n:=K_mK^{*n}+K^*_mK^n -\delta_m{}^n~.
\end{align}
Moreover every tensor can be decomposed into directions along and 
perpendicular to the $K$-orthogonal subspaces, using the 
projectors  
\beal
(\Pi^{\parallel})_m^n&:=\delta_m^n -\frac{1}{2}(K_mK^{*n}+K^{*}_mK^{ n}  )  \nn\\
(\Pi^{\perp})_m^n&:=\frac{1}{2}(K_mK^{*n}+K^{*}_mK^{ n}  )  ~.
\label{b20}
\end{align}
In particular, the metric decomposes as
\beal
g_{mn}=G_{mn}+\frac{1}{2}(K_mK^*_n+K^*_mK_n)~,
\label{uop}
\end{align}
where $K^mG_{mn}=0$. Finally, note the useful identity
\beal
\varepsilon^{\parallel}_{mnpqrs}=-15\widehat{J}_{[mn}\widehat{J}_{pq}\widehat{J}_{rs]}~,
\label{b0}
\end{align}
which can be proven by contracting 
the last line of (\ref{jids}) with $K^tK^{* u}$.

\subsection*{(Anti)holomorphic six-cycles}
\label{tut}

We shall be interested in particular in the case where 
the almost product structure on $X$, defined in the previous 
section, is integrable and 
the metric on $X$ can be brought to the standard form of a 
fibration over a six-cycle ${\Sigma}$:
\beal
ds^2(X)=G_{mn}(x)dx^m\otimes dx^n+(dz+A)\otimes(dz^*+A^*)~.
\label{explfibr}
\end{align}
This is of the form (\ref{uop}), where 
$K^*=dz+A(x)$ is the one-form dual of the 
holomorphic Killing vector $\partial/\partial z$, 
the $x^m$s are the coordinates on the base, $z$ is  
a complex coordinate on the normal fibre and $A(x)$ 
is a complex connection one-form. The six-cycle defined by the fibration is a holomorphic cycle, 
with a similar definition for an antiholomorphic cycle.   
We would like to stress that in general $X$ is not the total space 
of the normal bundle over ${\Sigma}$, but this approximation 
becomes more accurate as the size of ${\Sigma}$ is scaled up. 

Note that for any $S_{m}$ such that 
$S_{m}=(\Pi^\perp)_{m}^qS_{q}$,  we have 
\beal
\nabla^{\parallel}_mS^{m}&=(\Pi^\parallel)_m^q\nabla_q(\Pi^\perp)_n^mS^{n}
=\frac{1}{2}(\Pi^\parallel)^{mq}(K_n\nabla_qK^*_m+{\rm c.c.})S^{n}=0~.
\label{tba}
\end{align}
In the first equality we have used the orthogonality of $\Pi^{\perp}$, $\Pi^{\parallel}$. 
In the second equality we have taken (\ref{b20}) into account, and we have 
noted that $K^m(\Pi^\parallel)_m^n=0$. 
In the last equality we have taken into account that $(\Pi^\parallel)^{mn}$ is symmetric, and that 
$K$ is Killing. Similarly we can prove that if $S_{m}=(\Pi^\parallel)_{m}^qS_{q}$,  we have 
\beal
\nabla^{\perp}_mS^{m}&=0~.
\label{tbb}
\end{align}
Equations (\ref{tba},\ref{tbb}) can also be generalized to $p$-forms.



\section{Gravitino vertex operator}
\label{pstapp}

In this appendix we give the details of the derivation of the gravitino vertex operator 
of section \ref{grvsec}, equation (\ref{grv}). 

For any $Q$, let $\D Q$ the part of $Q$ linear in the gravitino. 
From (\ref{action}) and the analysis of the previous sections we find
\bead
g_{mn}&=g^{(0)}_{mn}+g^{(1)}_{mn}+g^{(2)}_{mn}+{\cal O}(\Psi,\th^3)\nn\\
\D g_{mn}&=\D g^{(1)} _{mn}+\D g^{(2)}_{mn}+\D g^{(3)}_{mn}
+{\cal O}(\Psi^2, \th^5)~,
\end{alignat}
where
\bead
g^{(0)}_{mn}&=G_{mn}\nn\\
g^{(1)}_{mn}&=-i(\cD_{(m}\th\C_{n)}\th)\nn\\
g^{(2)}_{mn}&=-\frac{1}{4}(\cD_m\th\C^{\una}\th)(\cD_n\th\C_{\una}\th)\nn\\
\D g^{(1)}_{mn}&=-2i(\Psi_{(m}\C_{n)}\th)\nn\\
\D g^{(2)}_{mn}&=-(\Psi_{(m}\C^{\una}\th)(\cD_{n)}\th\C_{\una}\th)\nn\\
\D g^{(3)}_{mn}&=\frac{1}{3}(\Psi_{(m}\fg\C_{n)}\th)
+\frac{1}{3}(\Psi_{\unp\unq}{\cal I}_{(m}{}^{\unp\unq}\C_{n)}\th)
\end{alignat}
For the inverse of the Green-Schwarz metric we have
\bead
g^{mn}&=g_{(0)}^{mn}+g_{(1)}^{mn}+g_{(2)}^{mn}+{\cal O}(\Psi,\th^3)\nn\\
\D g^{mn}&=\D g_{(1)}^{mn}+\D g_{(2)}^{mn}+\D g_{(3)}^{mn}
+{\cal O}(\Psi^2, \th^5)~,
\end{alignat}
where
\bead
g_{(0)}^{mn}&=G^{mn}\nn\\
g_{(1)}^{mn}&=i(\cD^{(m}\th\C^{n)}\th)\nn\\
g_{(2)}^{mn}&=\frac{1}{4}(\cD^m\th\C^{\una}\th)(\cD^n\th\C_{\una}\th)
-\frac{1}{4}(\cD^m\th\C^{p}\th)(\cD^n\th\C_{p}\th)\nn\\
&-\frac{1}{4}(\cD_p\th\C^{m}\th)(\cD^p\th\C^{n}\th)
-\frac{1}{2}(\cD_p\th\C^{(m}\th)(\cD^{n)}\th\C^{p}\th)\nn\\
\D g_{(1)}^{mn}&=-G^{mk}\D g^{(1)}_{kl} G^{ln}\nn\\
\D g_{(2)}^{mn}&=-G^{mk}\D g^{(2)}_{kl} G^{ln}-2g_{(1)}^{k(m}\D g^{(1)}_{kl} G^{n)l} \nn\\
\D g_{(3)}^{mn}&=-G^{mk}\D g^{(3)}_{kl} G^{ln}-2g_{(1)}^{k(m}\D g^{(2)}_{kl} G^{n)l}\nn\\
&-2g_{(2)}^{k(m}\D g^{(1)}_{kl}G^{n)l}-g_{(1)}^{mk}\D g^{(1)}_{kl} g_{(1)}^{ln}
~.
\end{alignat}
For later use note that
\bead
\frac{1}{\sqrt{-g}}=\Big(\frac{1}{\sqrt{-g}}\Big)^{(0)}
+\Big(\frac{1}{\sqrt{-g}}\Big)^{(1)}+\Big(\frac{1}{\sqrt{-g}}\Big)^{(2)}+{\cal O}(\Psi,\th^3)~,
\end{alignat}
where
\bead
\Big(\frac{1}{\sqrt{-g}}\Big)^{(0)}&=\frac{1}{\sqrt{-G}}\nn\\
\Big(\frac{1}{\sqrt{-g}}\Big)^{(1)}&=\frac{i}{2\sqrt{-G}}(\cD_m\th\C^m\th)\nn\\
\Big(\frac{1}{\sqrt{-g}}\Big)^{(2)}&=-\frac{1}{8\sqrt{-G}}\Big\{
(\cD_m\th\C^m\th)^2-(\cD_m\th\C^{\una}\th)(\cD^m\th\C_{\una}\th)\nn\\
&~~~~~~~~~~~~~~+(\cD_m\th\C^{p}\th)(\cD^m\th\C_{p}\th)+(\cD_m\th\C^{p}\th)(\cD_p\th\C^{m}\th)
\Big\}
~.
\end{alignat}
Moreover
\bead
H_{mnp}&=H^{(0)}_{mnp}+H^{(1)}_{mnp}+H^{(2)}_{mnp}+{\cal O}(\Psi,\th^3)\nn\\
\D H_{mnp}&= \D H^{(1)}_{mnp}+\D H^{(2)}_{mnp}+\D H^{(3)}_{mnp}+{\cal O}(\Psi^2, \th^5)~,
\end{alignat}
where
\bead
H^{(0)}_{mnp}&=F_{mnp}-c_{mnp}\nn\\
{H}_{mnp}^{(1)}&=\frac{3i}{2}(\cD_{[m}\th\C_{np]}\th)\nn\\
{H}_{mnp}^{(2)}&=\frac{3}{4}(\cD_{[m}\th\C_{n}{}^{\una}\th)(\cD_{p]}\th\C_{\una}\th)\nn\\
\D H^{(1)}_{mnp}&=3i(\Psi_{[m}\C_{np]}\th)\nn\\
\D H^{(2)}_{mnp}&=(\Psi_{[m}\C_{n}{}^{\una}\th)(\cD_{p]}\th\C_{\una}\th )
+2(\Psi_{[m}\C^{\una}\th)(\cD_{n}\th\C_{p]}{}_{\una}\th)\nn\\
\D H^{(3)}_{mnp}&=-\frac{1}{2}(\Psi_{[m}\fg\C_{np]}\th)
-\frac{1}{2}(\Psi_{\unp\unq}{\cal I}_{[m}{}^{\unp\unq}\C_{np]}\th)
~.
\end{alignat}
Similarly
\bead
v_{p}&=v^{(0)}_p+v^{(1)}_p+v^{(2)}_p+{\cal O}(\Psi,\th^3)\nn\\
\D v_{p}&=\D v^{(1)}_{p}+\D v^{(2)}_{p}+\D v^{(3)}_{p}+{\cal O}(\Psi^2, \th^5)~,
\end{alignat}
where
\bead
v^{(0)}_p&=a_p\nn\\
v^{(1)}_p&=\frac{i}{2}(\cD^m\th\C^n\th)a_ma_na_p  \nn\\
v^{(2)}_p&=\frac{1}{8}a_pa_ma_n\Big\{
(\cD^m\th\C_{\una}\th)(\cD^n\th\C^{\una}\th)
-(\cD^m\th\C_{p}\th)(\cD^n\th\C^{p}\th)-(\cD_p\th\C^{m}\th)(\cD^p\th\C^{n}\th)\nn\\
&~~~~~~~~~~~~~~~~~~~-2(\cD_p\th\C^{m}\th)(\cD^n\th\C^{p}\th)
-3(\cD^m\th\C^{n}\th)(\cD^q\th\C^{r}\th)a_qa_r
\Big\}\nn\\
\D v^{(1)}_{p}&=
\frac{1}{2}a_pa_ka_l\D g_{(1)}^{kl}\nn\\
\D v^{(2)}_{p}&=\frac{1}{2}v^{(1)}_p a_ka_l\D g_{(1)}^{kl}
+a_pa_kv^{(1)}_l\D g_{(1)}^{kl}+
\frac{1}{2}a_pa_ka_l\D g_{(2)}^{kl}\nn\\
\D v^{(3)}_{p}&=\frac{1}{2}v^{(2)}_p a_ka_l\D g_{(1)}^{kl}+a_pa_kv^{(2)}_l\D g_{(1)}^{kl}
+v^{(1)}_pv^{(1)}_ka_l\D g_{(1)}^{kl}+\frac{1}{2}
a_pv^{(1)}_kv^{(1)}_l\D g_{(1)}^{kl}\nn\\
&+\frac{1}{2}v^{(1)}_p a_ka_l\D g_{(2)}^{kl}
+a_pa_kv^{(1)}_l\D g_{(2)}^{kl}+
\frac{1}{2}a_pa_ka_l\D g_{(3)}^{kl}\nn\\
a_p&:=\frac{\partial_p a}{\sqrt{-G^{mn}\partial_ma\partial_na}}~.
\end{alignat}
Also
\bead
\widetilde{H}_{mn}&=\widetilde{H}_{mn}^{(0)}
+\widetilde{H}^{(1)}_{mn}+\widetilde{H}^{(2)}_{mn}+{\cal O}(\Psi,\th^3)\nn\\
\D \widetilde{H}_{mn}&=\D \widetilde{H}^{(1)}_{mn}+ \D \widetilde{H}^{(2)}_{mn}
+\D \widetilde{H}^{(3)}_{mn}+{\cal O}(\Psi^2, \th^5)~,
\end{alignat}
where
\bead
\widetilde{H}^{(k)}_{mn}&=\frac{1}{6}\epsilon^{m'n'pqrs}\Big(\frac{1}{\sqrt{-g}}g_{mm'}g_{nn'}
v_pH_{qrs}\Big)^{(k)}\nn\\
\D \widetilde{H}^{(k)}_{mn}&=\frac{1}{6}\epsilon^{m'n'pqrs}\sum^k_{i=1}
\Big\{\Big(\frac{1}{\sqrt{-g}}g_{mm'}g_{nn'}H_{qrs}\Big)^{(k-i)}
\D v^{(i)}_p \nn\\
&+\Big(\frac{1}{\sqrt{-g}}g_{mm'}g_{nn'}v_p \Big)^{(k-i)} \D H^{(i)}_{qrs}
-\frac{1}{2}\Big(\frac{1}{\sqrt{-g}}g_{mm'}g_{nn'}g^{kl}v_p H_{qrs} \Big)^{(k-i)} 
\D g^{(i)}_{kl} \nn\\
&+2\D g^{(i)}_{[m|m'}\Big(\frac{1}{\sqrt{-g}}g_{|n]n'}v_p H_{qrs}\Big)^{(k-i)}\Big\} 
~,~~~k=1,2,3~.
\end{alignat}
Finally, for the action we find
\bead
\D S&=T_{M5}\int_{{\Sigma}}{d^6x}
\sqrt{-G} \sum_{k=1}^{3}\Big\{\D {\cal L}^{(k)}_1+\D {\cal L}^{(k)}_2+\D {\cal L}^{(k)}_3
+{\cal O}(\Psi^2, \th^5)\Big\}
\end{alignat}
where
\bead
\D {\cal L}^{(k)}_1&=\frac{1}{2} \sqrt{
det(A_r{}^s) }\sum_{i=1}^{k}\Big(\Big\{1+\frac{1}{2}tr(A^{-1}B)+\frac{1}{8}\Big(tr(A^{-1}B)\Big)^2-\frac{1}{4}tr(A^{-1}B)^2 \Big\}\nn\\
&~~~~~~~~~\times(A^{-1}-A^{-1}BA^{-1}+A^{-1}BA^{-1}BA^{-1})^{mn}\Big)^{(k-i)}(\D g_{mn}-i\D\widetilde{H}_{mn})^{(i)}\nn\\ 
A_{mn}&:=G_{mn}+i\widetilde{H}^{(0)}_{mn}\nn\\
B_{mn}&:=(g_{mn}+i\widetilde{H}_{mn})^{(1)}+(g_{mn}+i\widetilde{H}_{mn})^{(2)}~.
\end{alignat}
Also
\bead
\D {\cal L}^{(k)}_2&= \frac{1}{4!\sqrt{-G}}
\epsilon^{mnpqrs} \sum_{i=1}^k \Big\{ \Big(H_{mnt}v_pv_lg^{tl}
\Big)^{(k-i)} \D H^{(i)}_{qrs} 
+\Big(H_{qrs}v_pv_lg^{tl}\Big)^{(k-i)}\D H^{(i)}_{mnt}
\nn\\
&+\Big(H_{qrs}H_{mnt}v_lg^{tl}\Big)^{(k-i)}\D v^{(i)}_p
+\Big(H_{qrs}H_{mnt}v_pg^{tl}\Big)^{(k-i)}\D v^{(i)}_l
+\Big(H_{qrs}H_{mnt}v_pv_l\Big)^{(k-i)}\D g_{(i)}^{tl}
\Big\}\nn\\
\D {\cal L}^{(k)}_3&= -\frac{1}{6!\sqrt{-G}}\epsilon^{mnpqrs} 
\Big(\D C^{(k)}_{mnpqrs}+10 F_{mnp}\D H^{(k)}_{qrs} \Big) ~,
\end{alignat}
where  
\bead
\Delta C^{(1)}_{m_1\dots m_6}&=-6i(\Psi_{[m_1}\C_{m_2\dots m_6]}\th) \nn\\
\Delta C^{(2)}_{m_1\dots m_6}&=
10(\Psi_{[m_1}\C^{\una}\th)(\cD_{m_2}\th\C_{m_3\dots m_6]\una}\th)
-5(\Psi_{[m_1}\C_{m_2\dots m_5| \una}\th)(\cD_{|m_6]}\th\C^{\una}\th)\nn\\
\Delta C^{(3)}_{m_1\dots m_6}&=(\Psi_{[m_1}\fg\C_{m_2\dots m_6]}\th)
+(\Psi_{\unp\unq}{\cal I}_{[m_1}{}^{\unp\unq}\C_{m_2\dots m_6]}\th)
~.
\end{alignat}

%
The gravitino vertex operator at order $\th^k$ is given by 
\bead
V=T_{M5}\int_{{\Sigma}}d^6x\sqrt{-G}\Big(\Psi^{\ua}_{\unm}V_{\ua}^{\unm}+\Psi^{\ua}_{\unm\unn}
V_{\ua}^{\unm\unn}\Big)
~, 
\end{alignat}
where
\bead
V_{\ua}^{\unm}&:=\frac{\partial}{\partial\Psi^{\ua}_{\unm}}\Big(
\D{\cal L}^{(k)}_1+\D  {\cal L}^{(k)}_2+\D {\cal L}^{(k)}_3\Big)\vert_{\Psi=0}\nn\\
V_{\ua}^{\unm\unn}&:=\frac{\partial}{\partial\Psi^{\ua}_{\unm\unn}}\Big(
\D{\cal L}^{(k)}_1+\D  {\cal L}^{(k)}_2+\D {\cal L}^{(k)}_3\Big)\vert_{\Psi=0}
\label{au}
\end{alignat}
%
%
and we have denoted by $\Psi^{\ua}_{\unm\unn}$ the gravitino field strength: 
$\Psi_{\unm\unn}:=\cD_{[\unm}\Psi_{\unn]}$. 
Substituting the preceding formul{\ae} in (\ref{au}) we arrive at the following explicit expressions
\subsection*{Order $\th$}
\bead
V_{~~~\ua}^{(1)\unm\unn}&=0\nn\\
V_{~~~\ua}^{(1)\unm}&=-i\sqrt{det(A_i{}^j)}(A^{-1})^{(mn)}(\C_{m}\th)_{\ua}\partial_{n}X^{\unm}\nn\\
&+\frac{\e^{lpqrs}{}_m}{6\sqrt{-G}}\sqrt{det(A_i{}^j)}(A^{-1})^{[mn]}
\Big\{
(\C_{n}\th)_{\ua}\partial_lX^{\unm}+(\C_{l}\th)_{\ua}\partial_nX^{\unm} \Big\} a_p (F_{qrs}-c_{qrs})\nn\\
&+\frac{\e^{klpqrs}}{12\sqrt{-G}}\sqrt{det(A_i{}^j)}(A^{-1})_{kl}
a_p \nn\\
&~~~~~~~~~~~~~~~~~\times\Big\{
(F_{qrs}-c_{qrs})\Big[ a^ma^n(\C_m\th)_{\ua}\partial_nX^{\unm} +(\C^{m}\th)_{\ua}\partial_mX^{\unm}
\Big] +3(\C_{qr}\th)_{\ua}\partial_sX^{\unm} \Big\} \nn\\
&+\frac{i\e^{klpqrs}}{12\sqrt{-G}}
a_k(F_{lpq}-c_{lpq})\Big\{(F_{rst}-c_{rst})
\Big[a^ta^n (\C_n\th)_{\ua}+\frac{1}{2}(\C^t\th)_{\ua} \Big]
+\frac{1}{2}(\C_{rs}\th)_{\ua}
\Big\} a^m\partial_mX^{\unm}\nn\\
&+\frac{i\e^{klpqrs}}{24\sqrt{-G}}
a_ka^n (F_{lpq}-c_{lpq}) (F_{rs}{}^{t}-c_{rs}{}^{t})
(\C_n\th)_{\ua}\partial_tX^{\unm}\nn\\
&+\frac{i\e^{klpqrs}}{5!\sqrt{-G}}
\Big\{  15  a^ta_k(F_{lpt}-c_{lpt})(\C_{qr}\th)_{\ua} 
-10 a^ta_k(F_{lpq}-c_{lpq})(\C_{rt}\th)_{\ua}\nn\\
&~~~~~~~~~~~~~~~~~~~~~~~~~~~~~~~~~~~~~~~~~~~~~~~~~~~~~~~~~~~~
- 5F_{klp}(\C_{qr}\th)_{\ua}  -(\C_{klpqr}\th)_{\ua} 
\Big\}\partial_s X^{\unm}
~.
\end{alignat}
\subsection*{Order $\th^2\cD\th$}
\bead
V_{~~~\ua}^{(2)\unm\unn}&=0\nn\\
V_{~~~\ua}^{(2)\unm}&=
-\Big\{ 
(\C^{\una}\theta)_{\ua}(\cD^m\th\C_{\una}\th)-2(\C_{n}\theta)_{\ua}(\cD^{(m}\th\C^{n)}\th)\nn\\
&~~~~~~~~~~-\frac{1}{6}(\C^m\C^{n\una}\theta)_{\ua}(\cD_n\th\C_{\una}\th)
+\frac{1}{6}(\C^m\C^{np}\theta)_{\ua}(\cD_n\th\C_{p}\th) \Big\}\partial_m X^{\unm}
+\dots
~,
\label{try}
\end{alignat}
where the ellipses stand for terms which drop out in the case of normal flux (see below). 
We have also dropped terms proportional to $\C^m\cD_m\theta$, which do not contain the zero mode. 
\subsection*{Order $\th^3$}
\bead
V^{(3)\unm\unn}_{~~~\ua}&= 
%
\frac{i}{6}({\cal I}_{\unp}{}^{\unm\unn})_{\ua}{}^{\ub}V_{~~~\ub}^{(1)\unp}\nn\\
V_{~~~\ua}^{(3)\unm}&= 
\frac{i}{6}\fg_{\ua}{}^{\ub}V_{~~~\ub}^{(1)\unm}
~.
\label{gravkoukou}
\end{alignat}
%
\subsection*{Normal  flux}

In the case of normal flux, {\it i.e.} 
when the world-volume two-form tensor is flat 
($F_{mnp}=0$) and  
the pullback of the three-form potential onto the fivebrane vanishes 
($c_{mnp}=0$), the previous expressions simplify considerably. 
In particular we have, 
\bead
iV_{~~~\ua}^{(1)\unm}&=(\C^{m}\th)_{\ua}\partial_{m}X^{\unm}
+\frac{\e^{klpqrs}}{5!\sqrt{-G}}
(\C_{klpqr}\th)_{\ua} 
\partial_s X^{\unm}\nn\\
&=2(\C^{m}\th)_{\ua}\partial_{m}X^{\unm}
~.
\end{alignat}
In deriving the second equality above  we have taken into account  that
\bead
\frac{\varepsilon^{m_1\dots m_p m_{p+1}\dots m_6} }{p!\sqrt{-G}}\C_{m_{p+1}\dots m_6}=
-(-1)^{ p(p-1)/2 }\C^{m_1\dots m_p}P^+~,
\label{gammahodge}
\end{alignat}
where the projector $P^+$ is defined in equation 
(\ref{iyu}), 
and we have noted that 
after gauge-fixing the physical 
fermion modes satisfy $\theta=P^+\theta$. 
Moreover, the terms of the form $\Psi\th^2\cD\th$ 
can be read off of equation (\ref{try}). 
Taking (\ref{gammahodge}, \ref{gravkoukou}) into account 
and Wick-rotating,  we finally arrive at equation (\ref{grv}).


\section{Notation/conventions}
\label{notation/conventions}

For the convenience of the reader we give here an index of our 
main conventions and notation.

$\unM=(\unm,\unmu)$: target-space bosonic, fermionic curved indices

$\unA=(\una,\ua)$: target-space bosonic, fermionic flat indices

$M=(m,\mu)$: world-volume bosonic, fermionic curved indices (from the 
beginning of the paper, up to and including section \ref{supersymmetriccycles})

$m, n, p,\dots$: bosonic indices along $X$ (from section \ref{zeromodes} to the 
end of the paper)

$A=(a,\a)$: world-volume bosonic, fermionic flat indices

$Z^{\unM}=(X^{\unm},\th^{\unmu})$: bosonic, fermionic superembedding coordinates

$e_{\unm}{}^{\una}$: the $\th=0$ component of $E_{\unm}{}^{\una}$

$\Psi_{\unm}{}^{\ua}$: the $\th=0$ component of $E_{\unm}{}^{\ua}$

$g_{mn}:= E_{m}{}^{\una} E_{n}{}^{\unb} \eta_{\una\unb}$

$G_{\unm\unn}:=e_{\unm}{}^{\una}e_{\unn}{}^{\unb}\eta_{\una\unb}$

$(\omega_{\unm})_{\ua}{}^{\ub}$: the spin connection compatible with $G_{\unm\unn}$

$c_{\unm\unn\unp}$: the $\th=0$ component of $C_{\unm\unn\unp}$

$G_{\unm\unn\unp\unq}:=4\partial_{[\unm}c_{\unn\unp\unq]}$

$(\cD_{\unm})^{\ua}{}_{\ub}:=\d^{\ua}{}_{\ub}\partial_{\unm}
-\frac{1}{4}(\omega_{\unm})^{\ua}{}_{\ub}+\frac{1}{36}\Big(
(\C^{\una\unb\unc})^{\ua}{}_{\ub}G_{\unm\una\unb\unc}
-\frac{1}{8}(\C_{\unm}{}^{\una\unb\unc\und})^{\ua}{}_{\ub}G_{\una\unb\unc\und}
\Big)$

$(\cR_{\unm\unn})^{\ua}{}_{\ub}$: 
the curvature of the supercovariant derivative $(\cD_{\unm})^{\ua}{}_{\ub}$

Convention: On $x$-space forms we convert between flat and curved 
indices using  $e_{\unm}{}^{\una}$.

Convention: We pull-back superforms ($x$-space forms) onto the world volume using 
$\partial_mZ^{\unM}$ ($\partial_mX^{\unm}$). Hence 
$e_m{}^{\una}:=\partial_mX^{\unm}e_{\unm}{}^{\una}$, 
but $E_m{}^{\una}:=\partial_mZ^{\unM}E_{\unM}{}^{\una}$.

Convention: We raise/lower curved world-volume indices on superforms using the Green-Schwarz metric 
$g_{mn}$

Convention: We raise/lower curved world-volume indices on $x$-space forms using the metric 
$G_{mn}$




\begin{thebibliography}{99}





\bibitem{reva}
  M.~Grana,
  ``Flux compactifications in string theory: A comprehensive review,''
  Phys.\ Rept.\  {\bf 423} (2006) 91, hep-th/0509003.



\bibitem{revb}
  M.~R.~Douglas and S.~Kachru,
  ``Flux compactification,'' hep-th/0610102.



\bibitem{revc}
  R.~Blumenhagen, B.~Kors, D.~Lust and S.~Stieberger,
  ``Four-dimensional string compactifications with D-branes, orientifolds and
  fluxes,'' hep-th/0610327.




\bibitem{kklt}
S.~Kachru, R.~Kallosh, A.~Linde and S.~P.~Trivedi,
``De Sitter vacua in string theory,''
Phys.\ Rev.\ D {\bf 68} (2003) 046005, hep-th/0301240.



\bibitem{w}
E.~Witten,
``Non-Perturbative Superpotentials In String Theory,''
 Nucl.\ Phys.\ B {\bf 474} (1996) 343, hep-th/9604030.







\bibitem{rs}
  D.~Robbins and S.~Sethi,
  ``A barren landscape,''
  Phys.\ Rev.\  D {\bf 71} (2005) 046008, hep-th/0405011.





\bibitem{ktt}
L.~Gorlich, S.~Kachru, P.~K.~Tripathy and S.~P.~Trivedi,
``Gaugino condensation and nonperturbative superpotentials in flux
compactifications,'' hep-th/0407130.


\bibitem{saul}
N.~Saulina, ``Topological constraints on stabilized flux vacua,'' 
Nucl.\ Phys.\ B {\bf 720} (2005) 203, hep-th/0503125.



\bibitem{kall}
R.~Kallosh, A.~K.~Kashani-Poor and A.~Tomasiello,
``Counting fermionic zero modes on M5 with fluxes,''
JHEP {\bf 0506} (2005) 069, hep-th/0503138.



\bibitem{bw}
  C.~Beasley and E.~Witten,
  ``New instanton effects in string theory,''
  JHEP {\bf 0602} (2006) 060, hep-th/0512039.




\bibitem{bbs}
K.~Becker, M.~Becker and A.~Strominger,
``Five-branes, membranes and nonperturbative string theory,''
Nucl.\ Phys.\ B {\bf 456} (1995) 130, hep-th/9507158.


\bibitem{hm}
J.~A.~Harvey and G.~W.~Moore,
``Superpotentials and membrane instantons,''
hep-th/9907026.


\bibitem{ovru}
  B.~A.~Ovrut, T.~Pantev and J.~Park,
  ``Small instanton transitions in heterotic M-theory,''
  JHEP {\bf 0005} (2000) 045, hep-th/0001133.


\bibitem{lima}
  E.~Lima, B.~A.~Ovrut, J.~Park and R.~Reinbacher,
  ``Non-perturbative superpotential from membrane instantons in heterotic
  M-theory,''
  Nucl.\ Phys.\ B {\bf 614} (2001) 117, hep-th/0101049.


\bibitem{angu}
  L.~Anguelova and K.~Zoubos,
  ``Five-brane instantons vs flux-induced gauging of isometries,''
  JHEP {\bf 0610}, (2006), 061, hep-th/0606271.



\bibitem{buch}
  E.~I.~Buchbinder,
  ``Derivative F-terms from heterotic M-theory five-brane instanton,'' hep-th/0611119.



\bibitem{t}
D.~Tsimpis,
``Curved 11D supergeometry,''
JHEP {\bf 0411} (2004) 087, hep-th/0407244.



\bibitem{norcor}
I.~N.~McArthur,
``Superspace Normal Coordinates,''
Class.\ Quant.\ Grav.\  {\bf 1}, 233 (1984).



\bibitem{nicolai}
A.~Dasgupta, H.~Nicolai and J.~Plefka,
``Vertex operators for the supermembrane,''
JHEP {\bf 0005} (2000) 007, hep-th/0003280.




\bibitem{wittflux}
  E.~Witten,
  ``On flux quantization in M-theory and the effective action,''
  J.\ Geom.\ Phys.\  {\bf 22} (1997) 1, hep-th/9609122.

\bibitem{wittseth}
  S.~Sethi, C.~Vafa and E.~Witten,
  ``Constraints on low-dimensional string compactifications,''
  Nucl.\ Phys.\ B {\bf 480} (1996) 213, hep-th/9606122.



\bibitem{hmm}
  D.~Freed, J.~A.~Harvey, R.~Minasian and G.~W.~Moore,
  ``Gravitational anomaly cancellation for M-theory fivebranes,''
  Adv.\ Theor.\ Math.\ Phys.\  {\bf 2} (1998) 601, hep-th/9803205.



\bibitem{haaca}
  M.~Haack and J.~Louis,
  ``Duality in heterotic vacua with four supercharges,''
  Nucl.\ Phys.\ B {\bf 575}, 107 (2000), hep-th/9912181].


\bibitem{haacb}
 M.~Haack and J.~Louis,
  ``M-theory compactified on Calabi-Yau fourfolds with background flux,''
  Phys.\ Lett.\ B {\bf 507}, 296 (2001), hep-th/0103068.



\bibitem{bhs}
M.~Berg, M.~Haack and H.~Samtleben,
 ``Calabi-Yau fourfolds with flux and supersymmetry breaking,''
  JHEP {\bf 0304} (2003) 046, hep-th/0212255.

\bibitem{whs}
  B.~de Wit, I.~Herger and H.~Samtleben,
  ``Gauged locally supersymmetric D = 3 nonlinear sigma models,''
  Nucl.\ Phys.\ B {\bf 671} (2003) 175, hep-th/0307006.



\bibitem{haack}
  M.~Haack, D.~Krefl, D.~Lust, A.~Van Proeyen and M.~Zagermann,
  ``Gaugino condensates and D-terms from D7-branes,'' hep-th/0609211.



\bibitem{gano}
 O.~J.~Ganor,
  ``A note on zeroes of superpotentials in F-theory,''
  Nucl.\ Phys.\ B {\bf 499} (1997) 55, hep-th/9612077.





\bibitem{poortomasiello}
A.~K.~Kashani-Poor and A.~Tomasiello,
``A stringy test of flux-induced isometry gauging,''
  Nucl.\ Phys.\ B {\bf 728} (2005) 135, hep-th/0505208.




\bibitem{mayra}
  P.~Mayr,
  ``Mirror symmetry, N = 1 superpotentials and tensionless strings on
  Calabi-Yau four-folds,''
  Nucl.\ Phys.\ B {\bf 494} (1997) 489, hep-th/9610162.


\bibitem{howea}
  P.~S.~Howe and E.~Sezgin,
  ``D = 11, p = 5,''
  Phys.\ Lett.\ B {\bf 394} (1997) 62, hep-th/9611008.

\bibitem{howeb}
  P.~S.~Howe, E.~Sezgin and P.~C.~West,
  ``Covariant field equations of the M-theory five-brane,''
  Phys.\ Lett.\ B {\bf 399} (1997) 49, hep-th/9702008.

\bibitem{pst}
  I.~A.~Bandos, K.~Lechner, A.~Nurmagambetov, P.~Pasti, D.~P.~Sorokin and M.~Tonin,
  ``Covariant action for the super-five-brane of M-theory,''
  Phys.\ Rev.\ Lett.\  {\bf 78} (1997) 4332, hep-th/9701149.

\bibitem{schw}
  M.~Aganagic, J.~Park, C.~Popescu and J.~H.~Schwarz,
  ``World-volume action of the M-theory five-brane,''
  Nucl.\ Phys.\ B {\bf 496} (1997) 191, hep-th/9701166.

\bibitem{equi}
  I.~A.~Bandos, K.~Lechner, A.~Nurmagambetov, P.~Pasti, D.~P.~Sorokin and M.~Tonin,
  ``On the equivalence of different formulations of the M theory  five-brane,''
  Phys.\ Lett.\  B {\bf 408} (1997) 135, hep-th/9703127.





\bibitem{wfive}
  E.~Witten,
  ``Five-brane effective action in M-theory,''
  J.\ Geom.\ Phys.\  {\bf 22} (1997) 103, hep-th/9610234.



\bibitem{beloa}
  D.~Belov and G.~W.~Moore,
  ``Holographic action for the self-dual field,'' hep-th/0605038.

\bibitem{belob}
  D.~M.~Belov and G.~W.~Moore,
  ``Type II actions from 11-dimensional Chern-Simons theories,'' hep-th/0611020.




\bibitem{hklt}
P.~S.~Howe, S.~F.~Kerstan, U.~Lindstrom and D.~Tsimpis,
``The deformed M2-brane,'' JHEP {\bf 0309} (2003) 013, hep-th/0307072.

\bibitem{dk}
J.~M.~Drummond and S.~F.~Kerstan,
``Kappa-symmetric deformations of M5-brane dynamics,''
JHEP {\bf 0506} (2005) 003, hep-th/0412149.


\bibitem{ttt}
D.~Tsimpis,
``11D supergravity at ${\cal O}(l^3)$,''
JHEP {\bf 0410} (2004) 046; hep-th/0407271.




\bibitem{tt}
P.~K.~Tripathy and S.~P.~Trivedi,
``D3 Brane Action and Fermion Zero Modes in Presence of Background Flux,''
hep-th/0503072.


\bibitem{bb}
K.~Becker and M.~Becker,
``M-Theory on Eight-Manifolds,''
Nucl.\ Phys.\ B {\bf 477} (1996) 155, hep-th/9605053.





\bibitem{pvw}
K.~Peeters, P.~Vanhove and A.~Westerberg,
``Towards complete string effective actions beyond leading order,''
Fortsch.\ Phys.\  {\bf 52} (2004) 630, hep-th/0312211.




\bibitem{beck}
 K.~Becker and M.~Becker,
  ``Supersymmetry breaking, M-theory and fluxes,''
  JHEP {\bf 0107} (2001) 038, hep-th/0107044.

























\bibitem{ms}
D.~Martelli and J.~Sparks,
``G-structures, fluxes and calibrations in M-theory,''
Phys.\ Rev.\ D {\bf 68} (2003) 085014, hep-th/0306225.


\bibitem{teight}
D.~Tsimpis,
  ``M-theory on eight-manifolds revisited: N = 1 supersymmetry and generalized
  Spin(7) structures,''
  JHEP {\bf 0604} (2006) 027, hep-th/0511047.




\bibitem{dira}
  R.~Kallosh and D.~Sorokin,
  ``Dirac action on M5 and M2 branes with bulk fluxes,''
  JHEP {\bf 0505}, (2005) 005, hep-th/0501081.





\bibitem{wittenold}
 I.~Affleck, J.~A.~Harvey and E.~Witten,
 ``Instantons And (Super)Symmetry Breaking In (2+1)-Dimensions,''
  Nucl.\ Phys.\ B {\bf 206} (1982) 413.





\bibitem{gh}
Griffiths \& Harris,
``Principles of Algebraic Geometry,''
Wiley Classics Library Edition 1994.
























\bibitem{gg}
  M.~B.~Green and M.~Gutperle,
 ``Effects of D-instantons,''
  Nucl.\ Phys.\ B {\bf 498} (1997) 195, hep-th/9701093; 
  ``D-particle bound states and the D-instanton measure,''
  JHEP {\bf 9801} (1998) 005, hep-th/9711107; 
  ``D-instanton partition functions,''
  Phys.\ Rev.\ D {\bf 58} (1998) 046007, hep-th/9804123; 
 ``D-instanton induced interactions on a D3-brane,''
  JHEP {\bf 0002} (2000) 014, hep-th/0002011; 




\bibitem{bill}
  M.~Billo, M.~Frau, I.~Pesando, F.~Fucito, A.~Lerda and A.~Liccardo,
  ``Classical gauge instantons from open strings,''
  JHEP {\bf 0302} (2003) 045, hep-th/0211250.



\bibitem{blum}
R.~Blumenhagen, M.~Cvetic and T.~Weigand,
  ``Spacetime instanton corrections in 4D string vacua - the seesaw mechanism
  for D-brane models,'' hep-th/0609191;  
N.~Akerblom, R.~Blumenhagen, D.~Lust, E.~Plauschinn and M.~Schmidt-Sommerfeld,
  ``Non-perturbative SQCD superpotentials from string instantons,'' hep-th/0612132.











\end{thebibliography}
\end{document}